\documentclass[fleqn,usenatbib]{mnras}

\usepackage{mathptmx}
\usepackage{xspace}

\usepackage[T1]{fontenc}
\usepackage{ae,aecompl}

\usepackage{graphicx}	
\usepackage{amsmath}	
\usepackage{amssymb}	
\usepackage{rotating}

\newcommand{\kms}{\,km\,s$^{-1}$}
\newcommand{\around}{$\sim$}

\newcommand{\Halpha}{H$\alpha$}
\newcommand{\Hbeta}{H$\beta$}

\newcommand{\HI}{H{\sc i}}
\newcommand{\HII}{H{\sc ii}}
\usepackage{xspace}
\newcommand{\OIII}{\ensuremath{\text{[O\sc iii]}}\xspace}

\newcommand{\NII}{\ensuremath{\text{[N\sc ii]}}\xspace}

\newcommand{\degrees}{$^{\circ}$}
\newcommand{\orcid}[2]{\href{http://orcid.org/#2}{#1}}

\title[Emission Line Maps of the LMC]{Emission Line Velocity, Metallicity and Extinction Maps of the Large Magellanic Cloud}

\author[Lah et al.]
{\orcid{Philip Lah}{0000-0001-6841-6553}$^{1,2}$\thanks{E-mail: pkl9469@nyu.edu}, 
\orcid{Matthew Colless}{0000-0001-9552-8075}$^{3,4}$, 
\orcid{Francesco D'Eugenio}{0000-0003-2388-8172}$^{5,6}$,
\orcid{Brent Groves}{0000-0002-9768-0246}$^{7,4}$
and \orcid{Joseph D.~Gelfand}{0000-0003-4679-1058}$^{1,2}$
\\
$^{1}$New York University Abu Dhabi, PO Box 129188, Abu Dhabi, United Arab Emirates\\
$^{2}$Center for Astrophysics and Space Science (CASS), New York University Abu Dhabi, PO Box 129188, Abu Dhabi, UAE \\
$^{3}$Research School of Astronomy and Astrophysics, Australian National University, Canberra, ACT 2611, Australia\\
$^{4}$ARC Centre of Excellence for All Sky Astrophysics in 3 Dimensions (ASTRO 3D), Australia\\
$^{5}$Kavli Institute for Cosmology, University of Cambridge, Madingley Road, Cambridge, CB3 0HA, UK\\
$^{6}$Cavendish Laboratory, University of Cambridge, 19 JJ Thomson Avenue, Cambridge, CB3 0HE, UK\\
$^{7}$International Centre for Radio Astronomy Research, The University of Western Australia, 35 Stirling Hwy, 6009 Crawley, WA, Australia\\
}

\date{Accepted XXX. Received YYY; in original form ZZZ}

\pubyear{2022}

\begin{document}
\label{firstpage}
\pagerange{\pageref{firstpage}--\pageref{lastpage}}
\maketitle


\begin{abstract}
We measure the properties of optical emission lines in multiple locations across the Large Magellanic Cloud (LMC) using the Australian National University 2.3-metre telescope and the WiFeS integral field spectrograph. From these measurements we interpolate maps of the gas phase metallicity, extinction, \Halpha\ radial velocity, and \Halpha\ velocity dispersion across the LMC. The LMC metallicity maps show a complex structure that cannot be explained by a simple radial gradient. The bright \HII\ region 30~Doradus stands out as a region of high extinction. The \Halpha\ and \HI\ gas radial velocities are mostly consistent except for a region to the south and east of the LMC centre. The \Halpha\ velocity dispersion is almost always higher than the \HI\ velocity dispersion, except in the region that shows the divergence in radial velocity, where the \HI\ velocity dispersion is greater than the \Halpha\ velocity dispersion. This suggests that the \HI\ gas is diverging from the stellar radial velocity, perhaps as a result of inflow or outflow of \HI\ gas.  {The study of dwarf galaxies like the LMC is important as they are the building blocks of larger galaxies like our own Milky Way. The maps provided in this work show details not accessible in the study of more distant dwarf galaxies.}  
\end{abstract}





\section{Introduction}

The Large Magellanic Cloud (LMC), one of the closest galaxies to the Milky Way, is a gas-rich, metal-poor, actively star-forming, irregular, dwarf galaxy, with a stellar mass of $2.7 \times 10^{9}\,$M$_{\odot}$ \citep{vandermarel06}. {The distance to the LMC is 49.97\,kpc from an eclipsing-binary study \citep{pietrzynski2013}.} The LMC is similar to many high-redshift galaxies in terms of morphology, star formation rate, stellar mass, metallicity, and kinematics; {see similar galaxies at $1.2 <$z$< 2.5$ in \citet{curti20}. The study of dwarf galaxies is particularly important as they are the building blocks of larger galaxies within the scenario of hierarchical merging \citep{white78}, and indeed the LMC will merge with the Milky Way in the future \citep{cautun19}.  However studying dwarf galaxies is difficult because they are faint. With the LMC we have a dwarf galaxy close enough observe in detail.} However, due to its large angular size, it is difficult to observe its global properties and directly compare to resolved surveys of more distant galaxies. We address this problem by making many individual measurements across the LMC and interpolating between them. 

The stellar metallicity of the LMC has previously been traced by many authors \citep{olszewski91,cole05,grocholski06,carrera08,pompeia08,cioni09,lapenna12,olsen11,piatti13,narloch22}. The most detailed work comes from \citet{choudhury16}, who create a map of the stellar metallicity of the LMC using the slope of the red giant branch stars as an indicator of the local average metallicity. They found the LMC bar has an average metallicity of $-0.38 \pm 0.08$\,dex and a shallow gradient across the disc of $0.0578 \pm 0.0003$~dex~kpc$^{-1}$. The outer regions have lower metallicity, around $-$0.44\,dex.  {In comparison, the Milky Way disk has a metallicity around solar \citep[i.e.\ \around 0 dex;][]{casagrande11} and the Small Magellic Cloud has a metallicity around $-$0.94 dex \citep{choudhury18}.}

The emission line properties of \HII\ regions in the LMC have been determined by many authors including \citet{peimbert74}, \citet{dufour75}, \citet{pagel78}, \citet{dufour82}, \citet{garnett95}, \citet{tsamis03}, \citet{peimbert03}, \citet{pellegrini12}, \citet{selier12}, \citet{toribiosancipriano17}, \citet{mcleod19}, \citet{barman22}, \citet{crowther23} and \citet{jin23}. These observations, often at high spectral resolution, only cover a few \HII\ regions each. While, together, these studies have measured the gas phase metallicity in many regions of the LMC, they do not form a homogeneous set of observations and methods. These differences, and the large systematics underlying measurements of the gas metallicity \citep{kewley08}, make it difficult to reliably combine existing measurements and study the global gas phase metallicity properties of the LMC. In contrast, our observations sample locations across the entire LMC and use a single method to determine the gas phase metallicity.

{Of particular note in the study of the ionised gas in the LMC is the UM/CTIO Magellanic Cloud Emission-line Survey that used narrow-band filters at \OIII, \Halpha\ and [S{\sc ii}] to investigate the properties of the interstellar medium of the galaxy with resolution of 5~arcsec or better \citep{smith99}.  They created detailed maps of the \HII\ regions including many shells and super bubles. }

Dust reddening maps across the LMC have been made by many authors, including \citet{harris97}, \citet{hutchings01}, \citet{gordon03}, \citet{zaritsky04}, \citet{subramaniam05}, \citet{cox06}, \citet{dobashi08}, \citet{haschke11}, \citet{choi18}, \citet{joshi19}, \citet{demarchi21}, \citet{skowron21}, \cite{bell22} and \citet{chen22}. These maps have predominantly been made from stellar observations, such as the characteristic colour of red clump stars, and span all wavelengths from ultraviolet to infrared. Here we use the flux ratio of the \Halpha\ and \Hbeta\ emission lines (a method that has not been applied before to the LMC) to create a reddening map of the LMC, focussing on the \HII\ regions, that can provide an independent check and new insights.

Understanding the velocity structure of the LMC can give insights on the properties of dwarf galaxies, which form the majority of the galaxy population. The LMC's velocity structure has already been well studied.  \citet{alves00}, \citet{graff00}, and  \citet{grocholski06} use observations of stars to map the velocity structure, while \citet{reid06} used observations of planetary nebulae. \citet{ambrociocruz16}, using a scanning Fabry–Perot interferometer, measured the \Halpha\ emission line across almost the whole extent of the LMC, generating a photometric and kinematic catalogue of \HII\ regions and nebulae in the LMC. Observations of the LMC by \citet{kim03} and \citet{staveleysmith03}, using the Australian Telescope Compact Array (ATCA) and the Parkes telescope, obtained velocity information from the neutral hydrogen (\HI) 21~cm emission. While detailed maps of the \HI\ radial velocity and velocity dispersion already exist for the LMC, here we make the first comparison of the kinematics of the hot \Halpha-emitting gas and the cold neutral hydrogen gas.

In this paper we present the first large, consistent survey of multiple optical emission-lines in the LMC (Section~\ref{Observations}). We use this data to map the gas-phase metallicity of the LMC (Section~\ref{Gas_Phase_Metallicity}).  We map the extinction across the LMC using the flux ratio of the \Halpha\ and \Hbeta\ emission lines (Section~\ref{Extinction}). We also map the radial velocity and velocity dispersion of ionised gas within the LMC, as measured from the \Halpha\ emission line, and compare to maps of the \HI\ gas radial velocity and velocity dispersion (Sections~\ref{Radial_Velocity} and~\ref{Velocity_Dispersion}). Although \citet{ambrociocruz16} provide more detailed \Halpha\ radial velocity and velocity dispersion measurements, we compare our measurements with the \HI\ measurements made by \citet{kim03} and \citet{staveleysmith03} (the data from \citet{ambrociocruz16} was not available for this analysis). Our findings are summarised in Section~\ref{Conclusion}.  This work is particularly relevant in the era of the Local Volume Mapper \citep{konidaris20} to act as a comparison to that survey.


\section{Observations}
\label{Observations}


\begin{figure*}
  \includegraphics[width=\columnwidth]{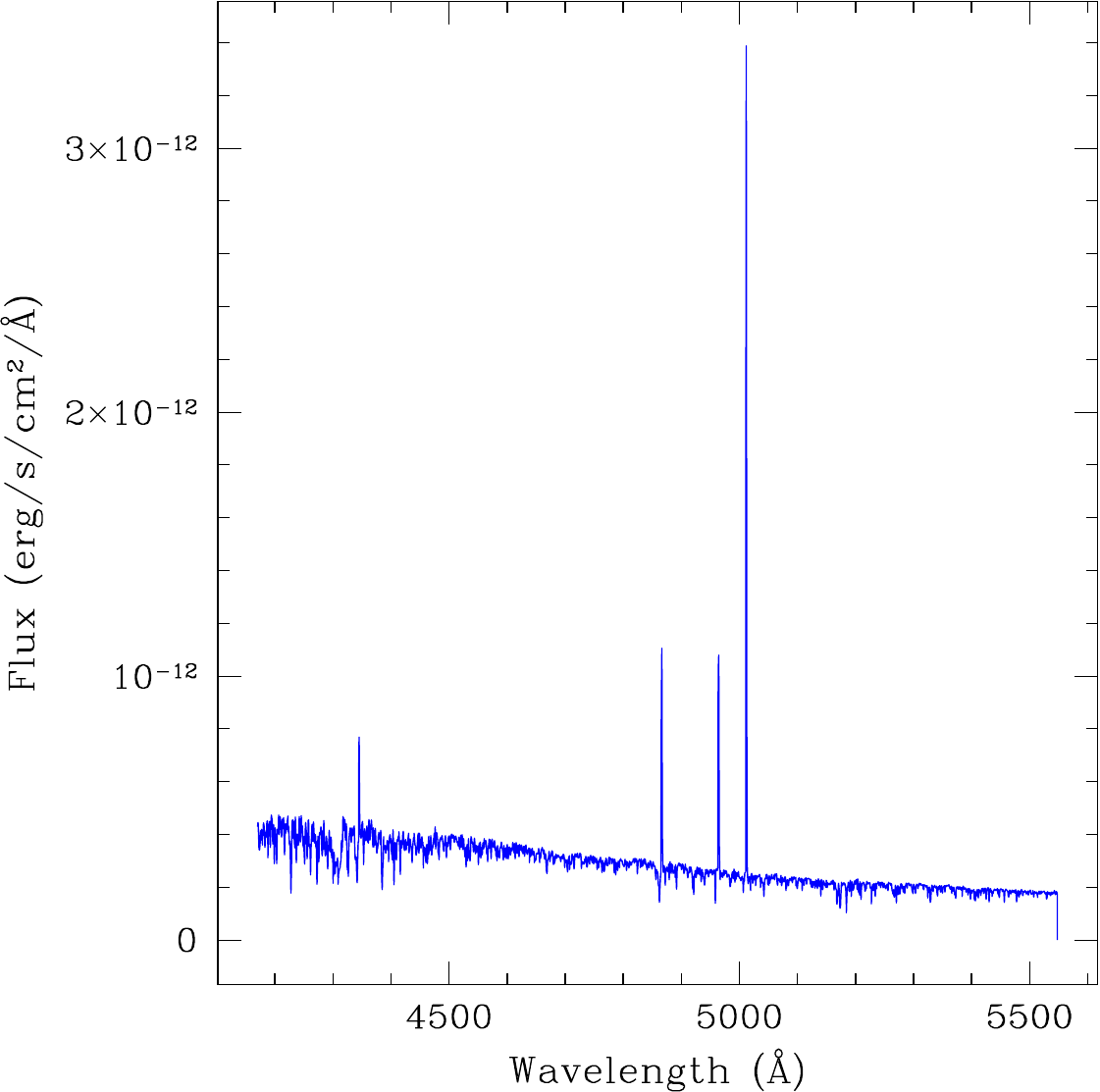}
  \includegraphics[width=\columnwidth]{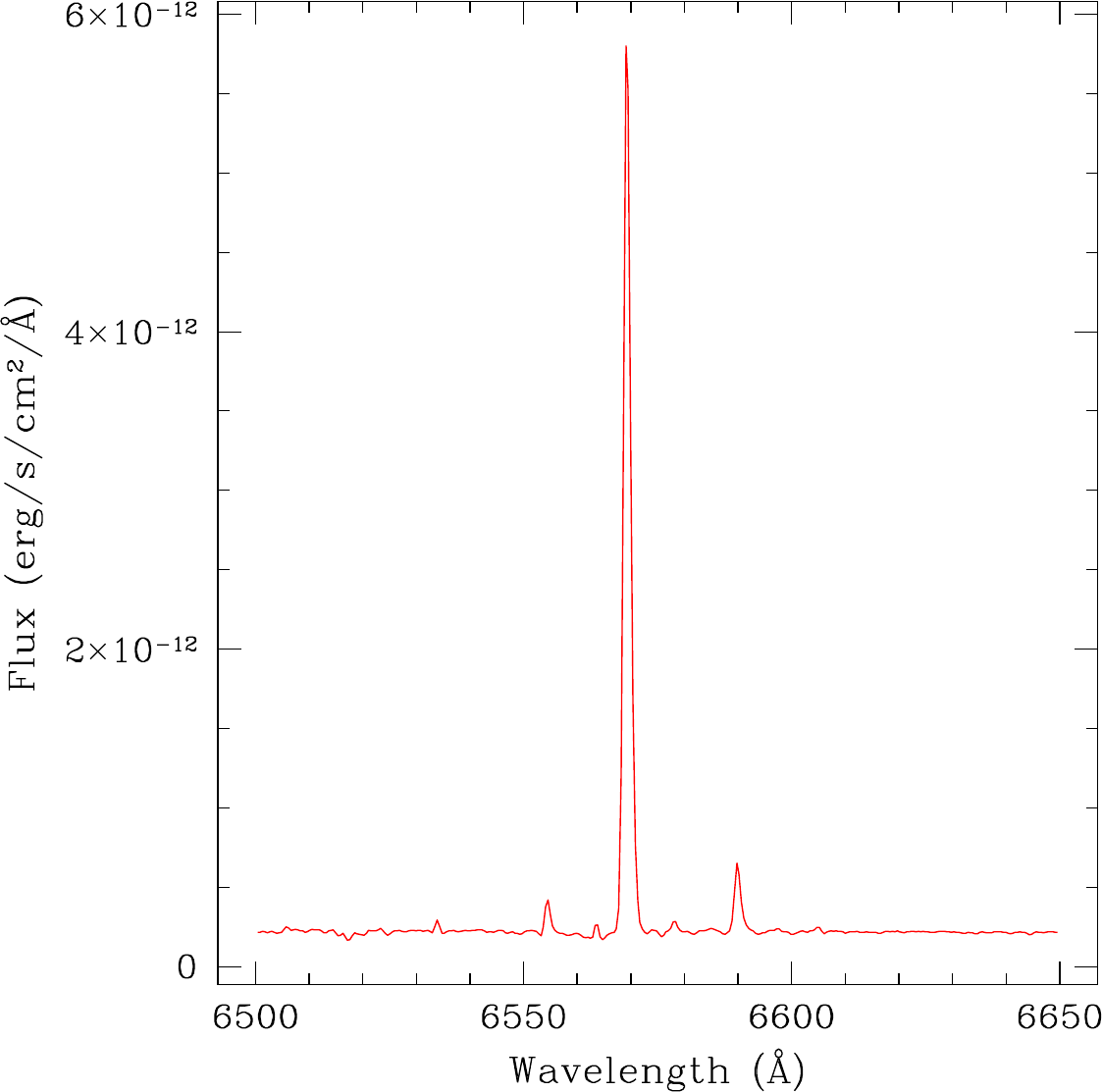}
  \caption{The panel on the left is an example blue spectrum. The continuum is dominated by moonlight. The panel on the right is an example red spectrum in the region of the \Halpha\ line.  The entire red spectrum is not shown, as it is dominated by sky lines. This spectrum comes from a field located at RA=05:35:00.151 and Dec=$-$69:12:16.69 (RA=83.751 and DEC=69.20).}
  \label{fig:plot_spectrum}
\end{figure*}


\begin{figure}
  \includegraphics[width=\columnwidth]{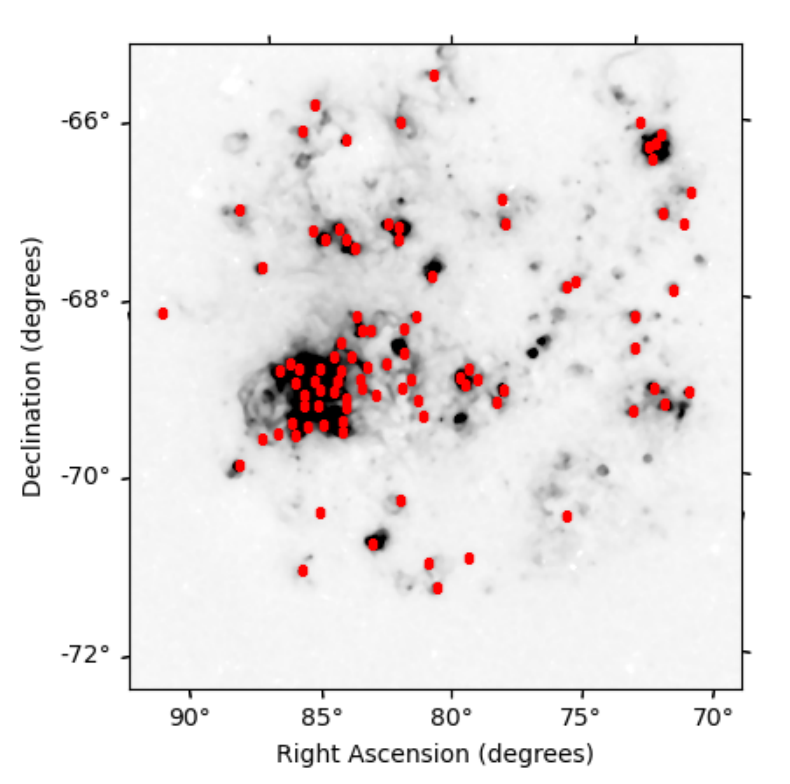}
  \caption{The positions of the 83 WiFeS fields used for this analysis, overlaid on the SHASSA H$\alpha$-emission image for the LMC. The field locations are shown by red dots; the coverage of the WiFeS IFU (37\,arcsec $\times$ 25\,arcsec) is considerably smaller than these dots and so does not usually cover the entire \HII\ region.  {Gaia parallaxes indicate the stellar disc of the LMC has an inclination angle of $52.7 \pm 2.4$~deg to the line of sight \citep{kadrmas21}. }} 
  \label{fig:LMC_final}
\end{figure}


\begin{figure}
  \includegraphics[width=\columnwidth]{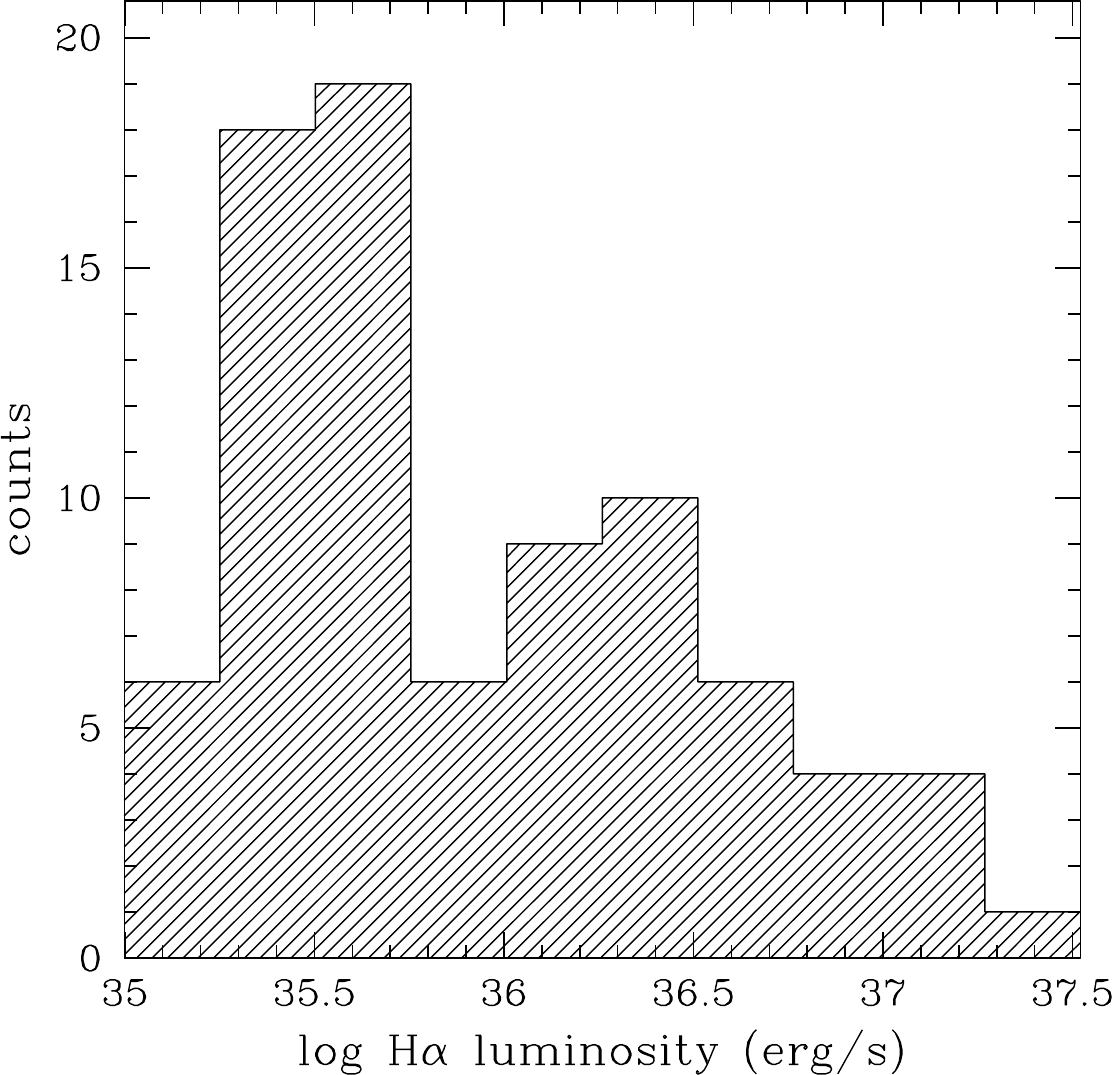}
  \caption{Histogram of \Halpha\ luminosity in the 83 WiFeS fields. The LMC distance is taken to be 49.97\,kpc \citep{pietrzynski2013}. The average error for the log \Halpha\ luminosity is 34.25 erg/s.}
  \label{fig:hist_luminosity_Halpha}
\end{figure}


The motivation for this project was to use bright time available on the Australian National University (ANU) 2.3-metre telescope with the WiFeS instrument \citep{dopita07,dopita10}. WiFeS is an integral field unit (IFU) spectrograph with a nominal field of view 38\,arcsec~$\times$~25\,arcsec and 0.5 arcsec sampling. This project was specifically designed to observe bright emission lines from the Large Magellanic Cloud that would still be visible when the moon was up.  {A targeted sample survey allowed us to spectroscopically observe \HII\ regions across the LMC in a fairly representative fashion.}

Targeted fields across the LMC were selected from the Southern H-Alpha Sky Survey Atlas (SHASSA) \citep{gaustad01}. SHASSA is a southern ($<+15$~degree Declination) survey using narrow (FWHM of 32\AA) and broad continuum filters centred at 6563~\AA\ (rest-frame \Halpha). The survey has a resolution of 1~arcimn. The sensitivity level of SHASSA is $1.2 \times 10^{-17}$~ergs~cm$^{-2}$~s$^{-1}$~arcsec$^{-2}$.  Based on this map, we selected a total of 116 fields to observe with WiFeS.  The fields were primarily chosen to lie on bright \HII\ regions in the SHASSA data, though a few additional fields were chosen lying between \HII\ regions, to try to obtain more regular sampling.  1200s exposures were used. For this project the signal is summed across the entire WiFeS IFU, giving a single spectrum that is \around 230 times brighter than the individual spaxel spectra. 

The fields across the LMC were observed using the B7000 ($R = 7000$) and R7000 ($R = 7000$) spectrographic gratings over 17 nights in the southern-hemisphere summer and autumn of 2021. The data was reduced using {\sc PyWiFeS} \citep{childress14}, using the observed bias, flat, and arc images.  Flux calibrations were done using observations of the standard star EG21.  {\sc PyWiFeS} produced two data cubes: one with wavelength coverage 4170--5548~\AA\ (the blue spectrum) and the other 5400--7000~\AA\ (the red spectrum). The data cubes were then intergrated to give a single spectrum. Simple summing was sufficient, as there was no large variation in the pixel-to-pixel noise level. When summing across the IFU, the bottom two rows of pixels were found to have a high noise level and were removed. This leaves the IFU summing over 37\,arcsec $\times$ 25\,arcsec. Example blue and red spectra are shown in Figure~\ref{fig:plot_spectrum}. {No separate sky subtraction was performed, as we were only interested in the flux of the continuum lines above the background. Therefore the entire background continuum was subtracted when measuring the lines.}

{Preliminary} individual radial velocities were measured from the 1-dimensional spectra using the Manual and Automatic Redshifting software \citep[{\sc MARZ},][]{hinton16}. The radial velocities are primarily determined from the H$\beta$\ emission line, with confirming lines from the \OIII lines at 4959\,\AA\ and 5007\,\AA. Overall, 104 fields had radial velocities determined in {\sc MARZ}. {More precise radial velocities were determined by fitting the lines.} 

A Python code was written to measure the properties of the \Hbeta, \OIII, \Halpha\ and \NII$\lambda 6548,6584$ emission lines from the WiFeS spectra.  In this code the two gratings are modelled independently.  Before the fit, we upscale the formal noise vector as a way to take into account correlations between the pixels. We do so by fitting the continuum as a straight line $\pm 120\text{--}150$~\AA\ from the lines (and masking the emission lines) and then calculating the robust standard deviation of the residuals. The formal uncertainties are upscaled so that their new median in the fitted region is equal to the standard deviation of the residuals.  The data, with the upscaled noise vector, are then modelled as the sum of a linear continuum and Gaussian emission lines; the latter are integrated over each spectral pixel. Within each grating, \Hbeta, \OIII, \Halpha\ and \NII may have different redshift and line width. The two doublets have fixed line ratios as prescribed from atomic physics. For each grating, we therefore have eight free parameters: two for the linear continuum, three for the recombination line, and three for the collisionally excited doublet. We find the best-fit parameters using a simple linear regression algorithm.  This gives us the flux of the lines, the radial velocity of the lines, and the FWHM of the lines. The uncertainties on these quantities are estimated by bootstrapping the data one hundred times.

For these observations the LMC continuum stellar light of \around 22~mag~arcsec$^{-2}$ in B band and \around 21~mag~arcsec$^{-2}$ in V~band \citep{hardy78} is well below the sky brightness, which is at least 15 times brighter. Any Balmer absorption from stellar atmospheres is therefore below the noise level, which justifies our linear model of the continuum as pure sky background. Measurements were made of the bright sky-lines at 6300.304~\AA\ and 6863.955~\AA\ so they could be used to correct any offset in radial velocity of the \Halpha\ emission line and quantify the instrumental resolution.

A subset of 83 fields were selected that had good \Halpha\ emission measurements for radial velocities and velocity dispersions. To be selected, a measurement had to have a signal-to-noise in the line~$\ge 3$, \Halpha\ Gaussian FWHM $>0.5$\AA\ and $<2$\AA\ (to ensure the line was not a noise spike or a match to the continuum), and Gaussian FWHM error $<0.1$\AA. In addition, the instrumental resolution measured from the sky-lines at the wavelength of \Halpha\ had to have Gaussian FWHM $<2$\AA.  The fields meeting these criteria primarily lie on bright knots of \Halpha\ emission visible in the narrowband SHASSA image, as can be seen in Figure~\ref{fig:LMC_final}. Fields between the knots of strong H$\alpha$\ emission were observed in an attempt to obtain a more uniform sampling across the LMC, but none of these fields met the quality criteria (mostly they failed to have observable emission lines). The measurements related to the \Halpha\ line can be found in Table~\ref{tab:Halpha}.  The histogram of the \Halpha\ luminosity in the 83 selected fields is shown in Figure~\ref{fig:hist_luminosity_Halpha}.


\section{Results}
\label{Results}

\subsection{Gas Phase Metallicity}
\label{Gas_Phase_Metallicity}


\begin{figure}
  \includegraphics[width=\columnwidth]{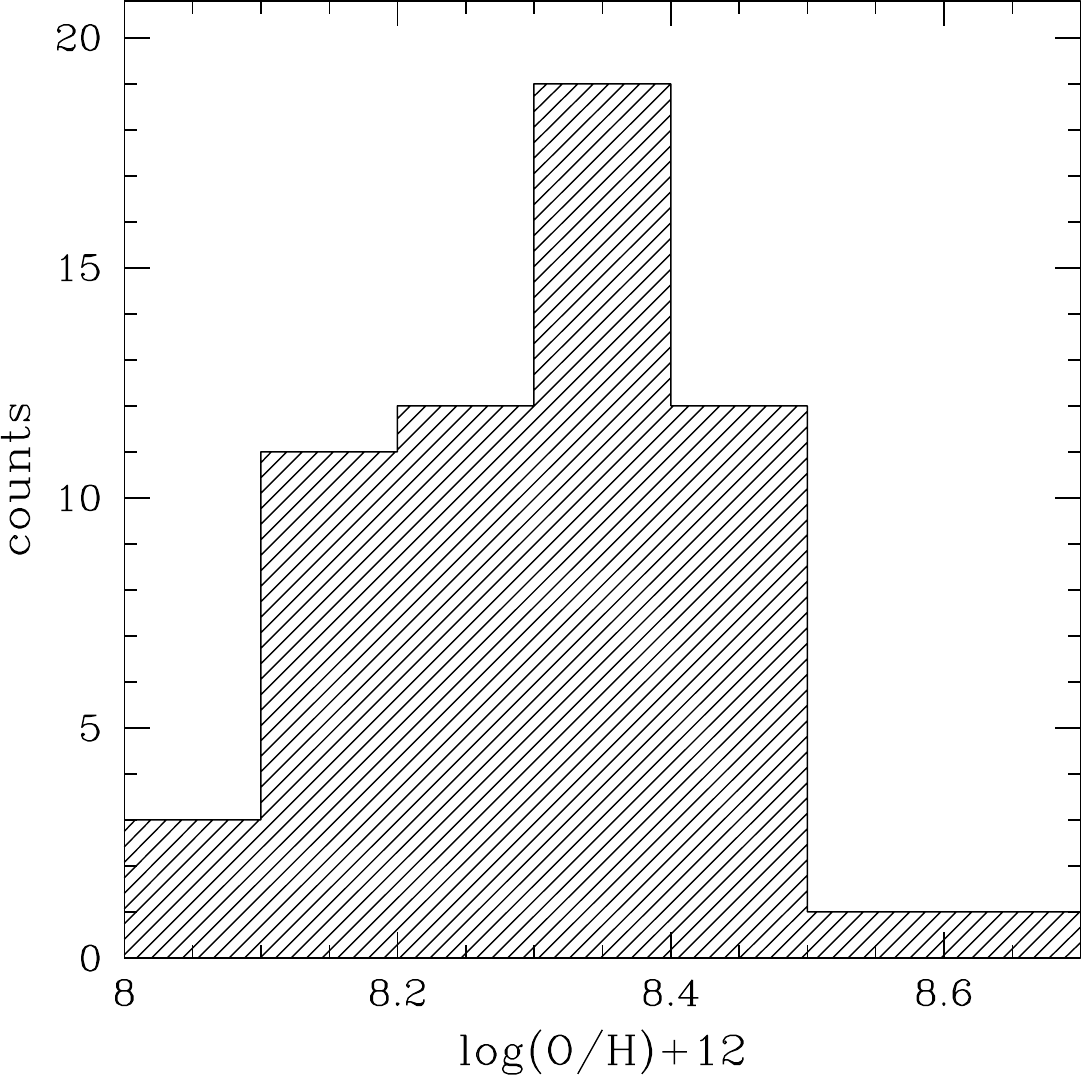}
  \caption{Histogram of the gas phase metallicity for each of the WiFeS fields included in this analysis.}
  \label{fig:hist_emission_lines_Kewely}
\end{figure}



\begin{figure}
  \includegraphics[width=0.99\columnwidth]{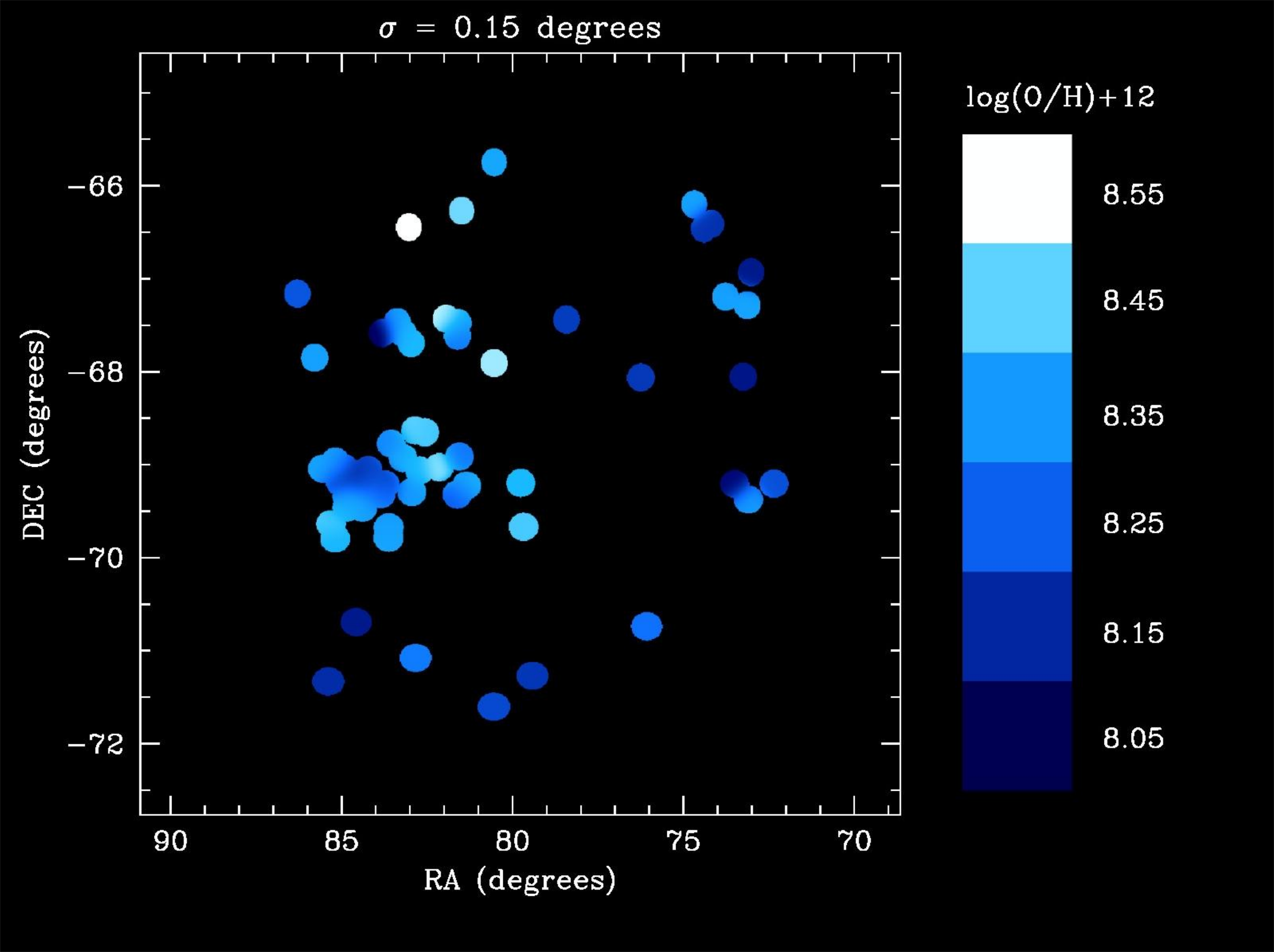}\\
  \includegraphics[width=0.99\columnwidth]{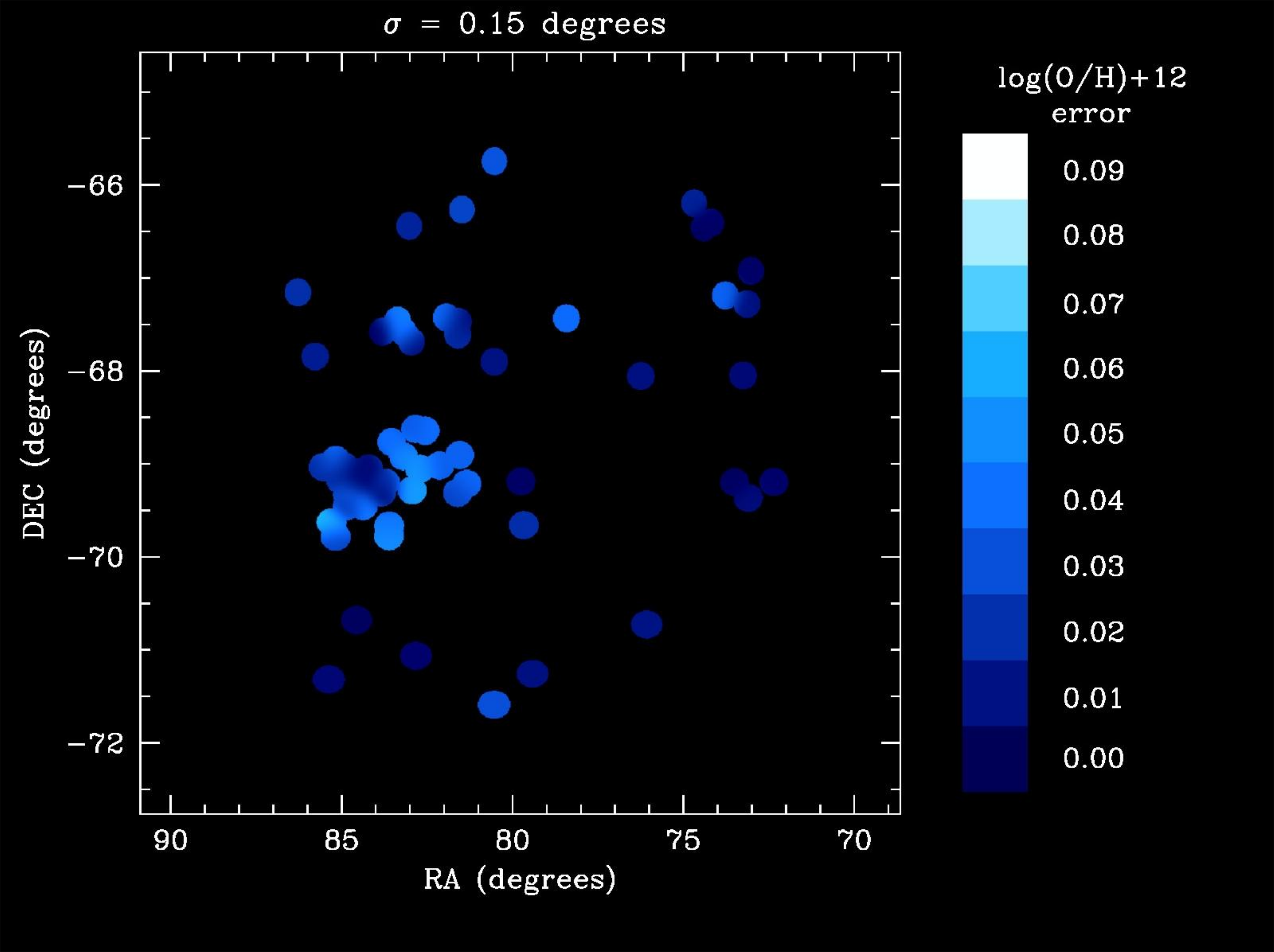}\\
  \includegraphics[width=0.99\columnwidth]{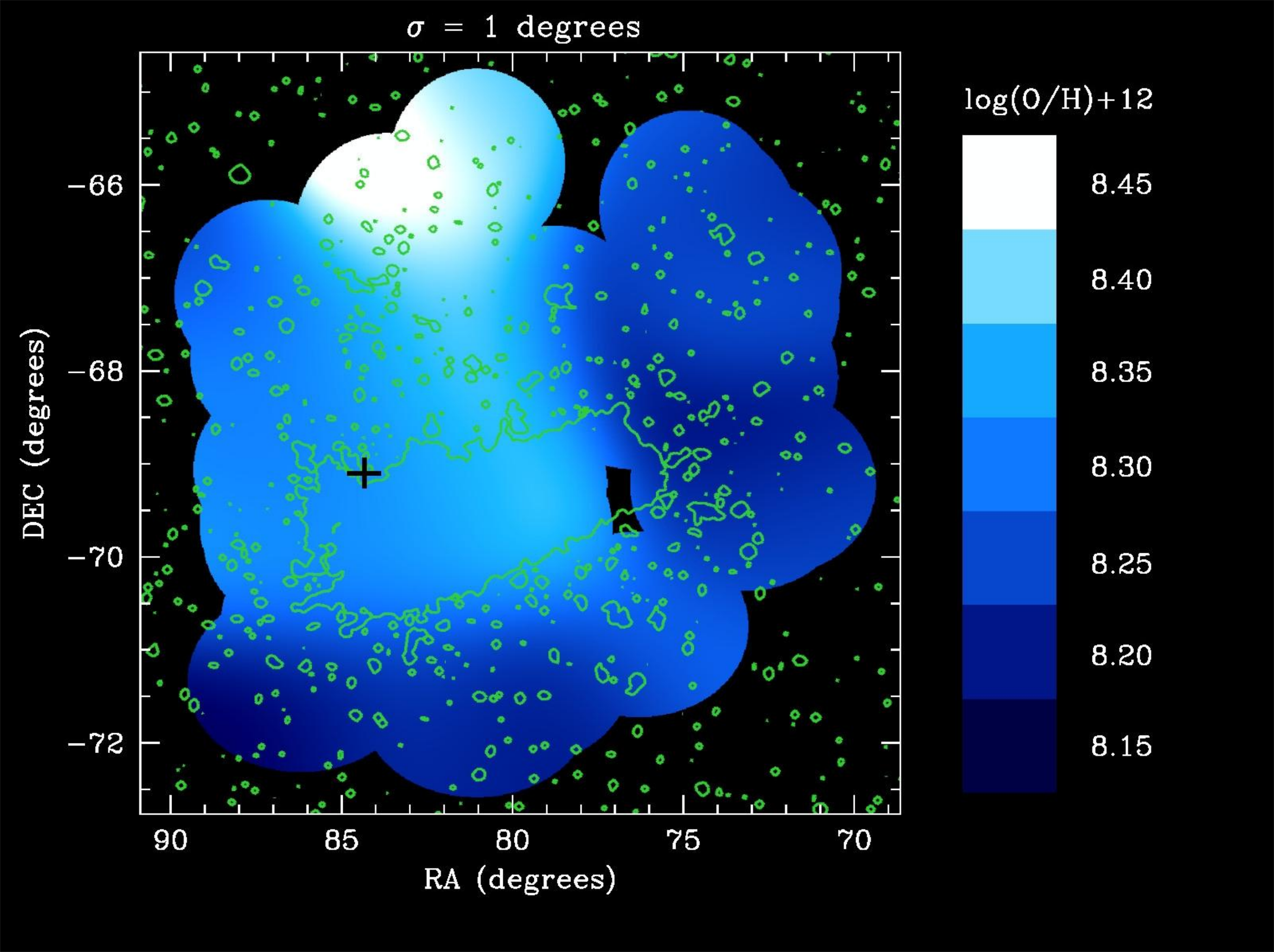}
  \caption{Interpolation of the gas phase metallicity using O3N2 across the LMC using a Gaussian kernel with $\sigma = 0.15$\,deg in the top panel and $\sigma = 1$\,deg in the bottom panel; the middle panel shows the error distribution for the metallicity. The interpolation has a cutoff distance from the nearest field equal to the $\sigma$ value. The contours in the bottom panel are derived from the SHASSA Continuum Image and highlight the LMC stellar bar.  The black cross in the bottom panel is the location of 30 Doradus.}
  \label{fig:metallicity}
\end{figure}


\begin{figure}
  \includegraphics[width=0.99\columnwidth]{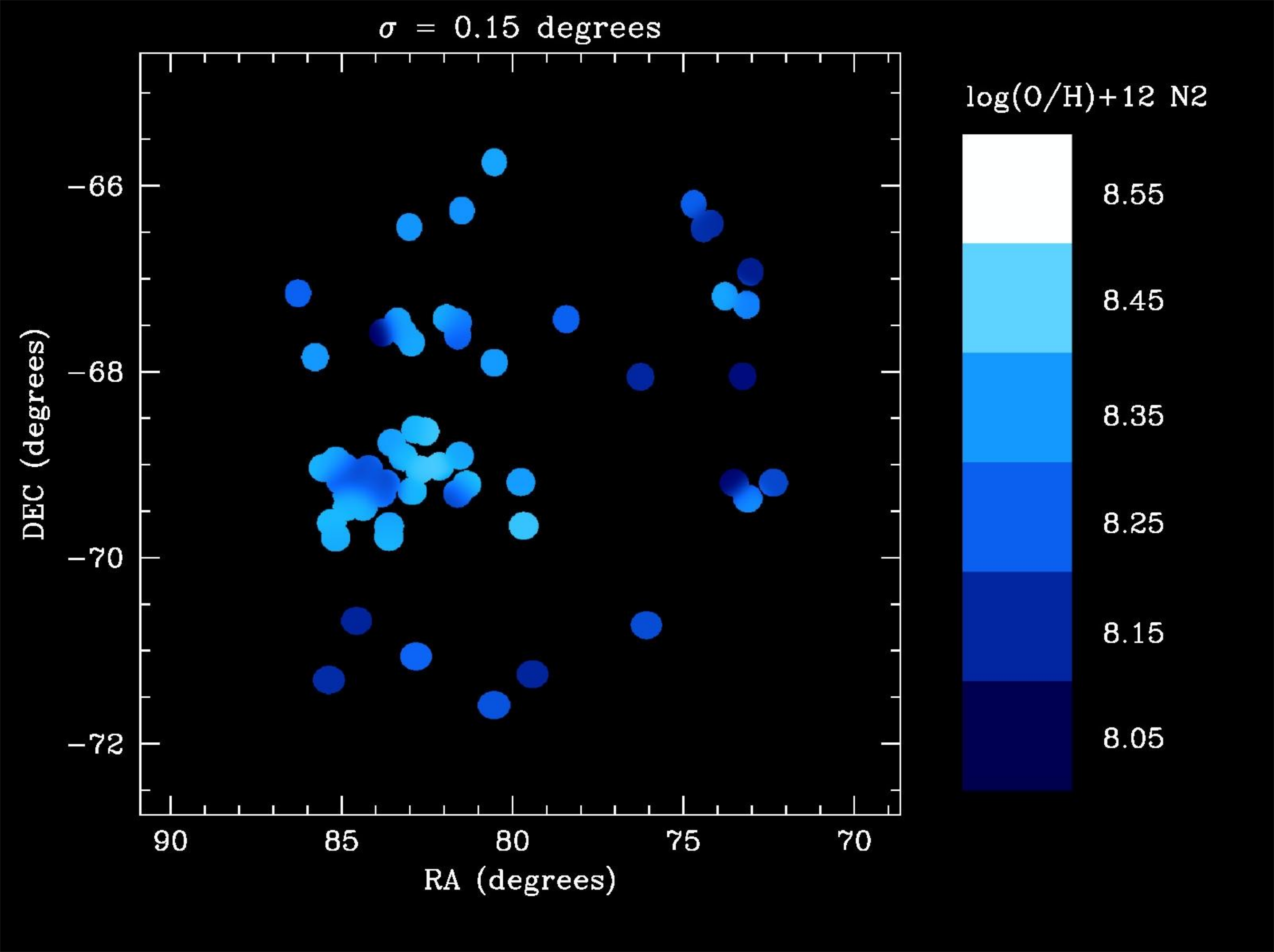}\\
  \includegraphics[width=0.99\columnwidth]{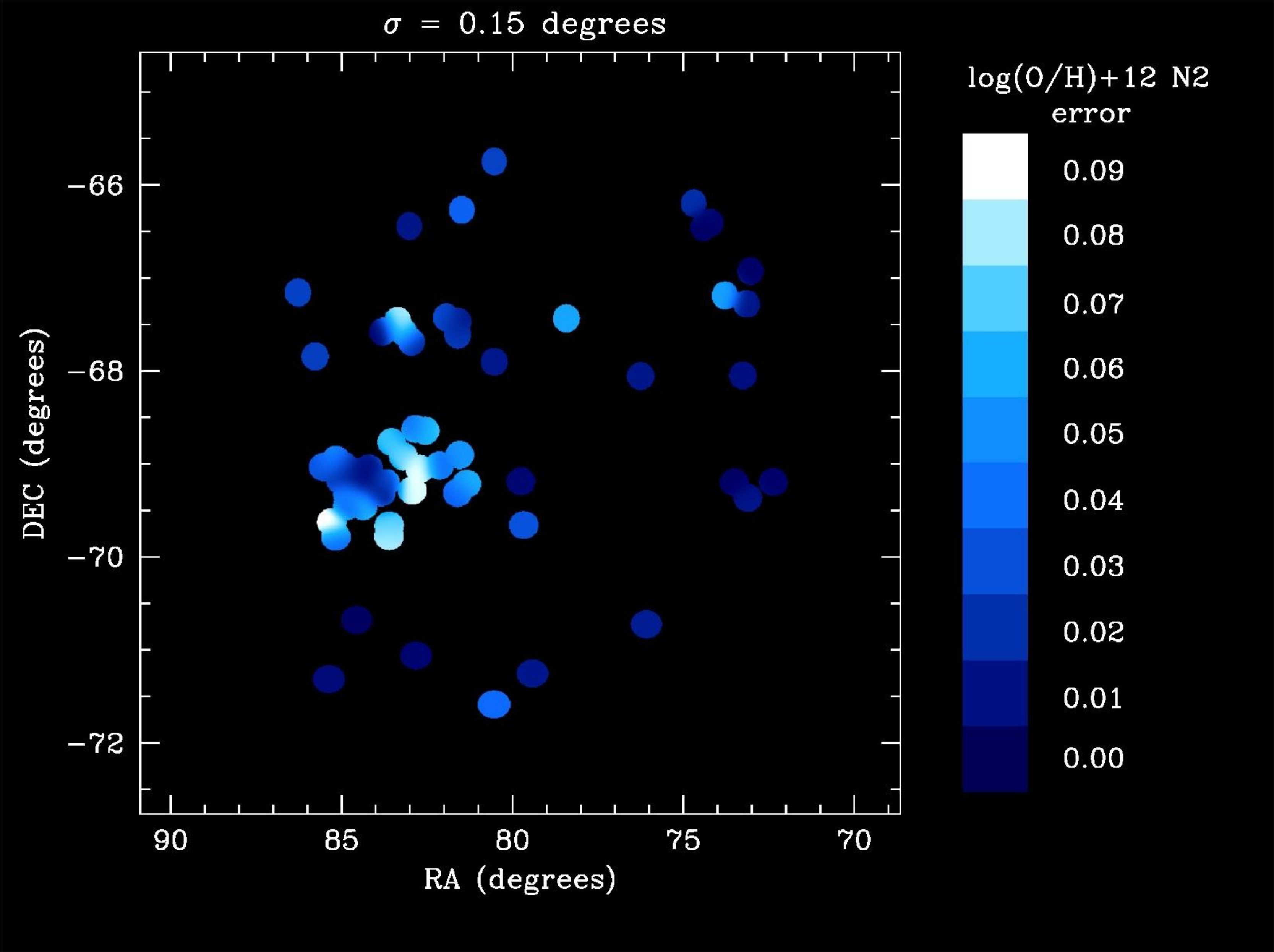}\\
  \includegraphics[width=0.99\columnwidth]{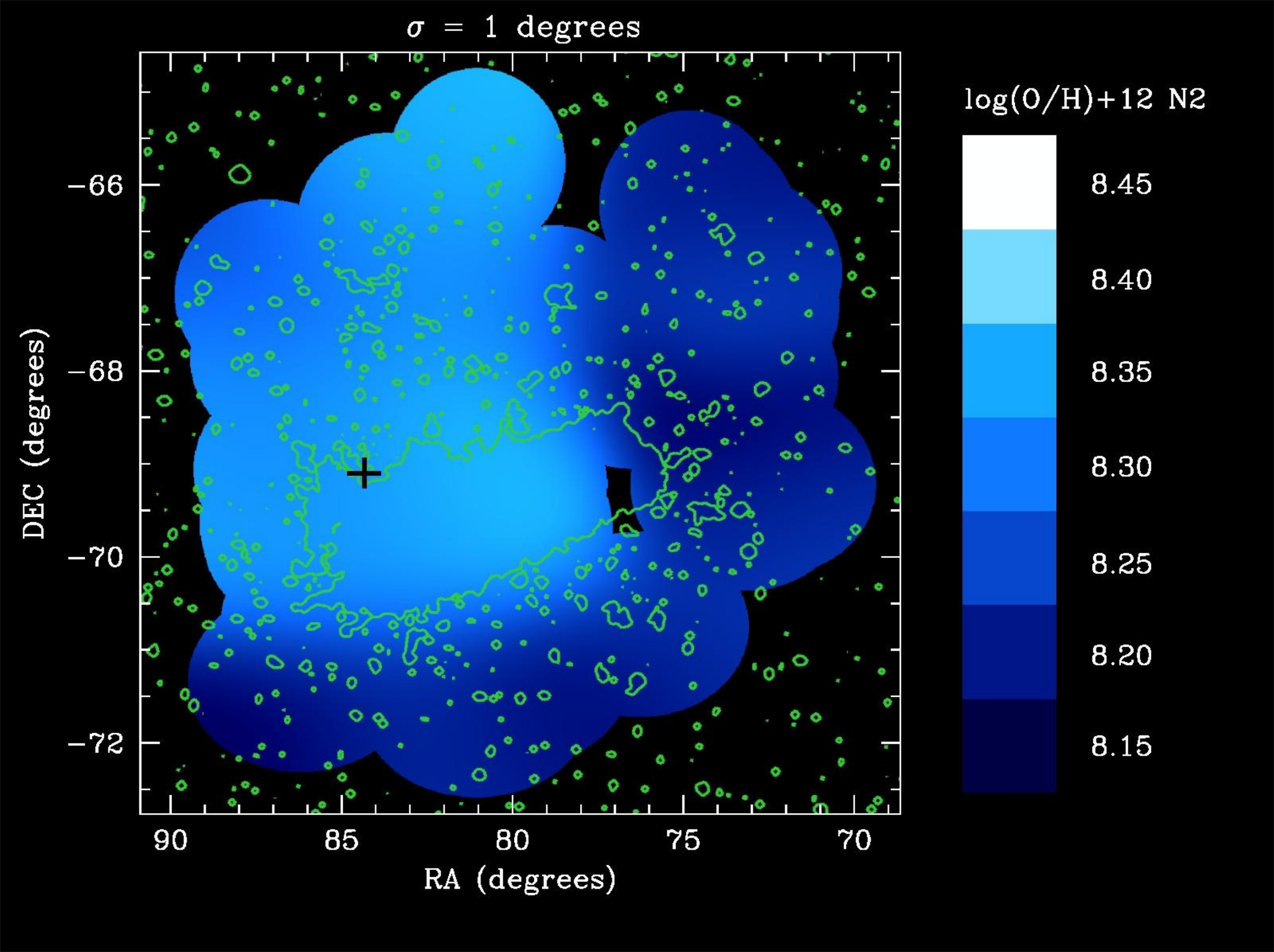}
  \caption{Interpolation of the gas phase metallicity using N2 across the LMC using a Gaussian kernel with $\sigma = 0.15$\,deg in the top panel and $\sigma = 1$\,deg in the bottom panel; the middle panel shows the error distribution for the N2 metallicity. The interpolation has a cutoff distance from the nearest field equal to the $\sigma$ value. The contours in the bottom panel are derived from the SHASSA Continuum Image and highlight the LMC stellar bar. The black cross in the bottom panel is the location of 30 Doradus.}
  \label{fig:metallicity_NII}
\end{figure}


A subsample of 59 fields were selected that had suitable H$\beta$\ \OIII~5007, H$\alpha$\ and \NII~6584 measurements for gas phase metallicity analysis. To be selected in this sample, each emission line measurement had to have a Gaussian FWHM $>0.5$\AA\ and $<2$\AA, and a signal-to-noise in the line~$\ge 3$. 

The gas phase metallicity was calculated for the LMC fields using the formula from \citet{pettini04}:
\begin{equation}
\rm 12 + \log(O/H) = 8.73 - 0.32 \times O3N2
\end{equation}
where
\begin{equation}
\rm O3N2 = \log \left( \frac{ [OIII]\lambda 5007/H\beta } {[NII]\lambda 6584/H\alpha} \right) ~.
\end{equation}
The advantage of this method of measuring the gas phase metallicity is that the emission lines in the ratio are both contained within either just the blue spectrum or just the red spectrum, so there is no error introduced by any differences in flux calibration between the blue and red spectra. Also, the lines in each ratio are close enough in wavelength that one can assume similar dust extinction for both lines, which then cancels from the ratio.  One caveat in using this method is that it only produces an estimate and depends heavily on the ionisation parameter of the nebula and its N/O ratio \citep{easeman23}.  The values for each field relevant to this work are found in Table \ref{tab:metallicity}. Figure~\ref{fig:hist_emission_lines_Kewely} is a histogram of the gas phase metallicity for the 59 selected WiFeS fields. The metallicities in these fields range from 8.06\,dex to 8.61\,dex, with a mean of 8.33\,dex (relatively low compared to most other galaxies; see \citet{tremonti04}). The median error in the metallicity of a field is 0.018\,dex; the largest error is 0.067\,dex.

To make a map of the metallicity in the LMC based on these 59 fields, we construct a regular grid of positions containing the LMC and calculate the gas phase metallicity for each point in this grid using a Gaussian kernel to interpolate between the measurements. That is, we sum over the metallicities in each of our fields using Gaussian weights determined by the distance between the field position and the grid position.  The formula for the weight, $w^2$, is
\begin{equation}
w^2 = \exp(-d^2/\sigma^2)
\end{equation}
where $d$ is the distance between the grid point and the field and $\sigma$ is the Gaussian kernel size. These weights multiply the metallicities for each field, which are then summed and normalised to get the metallicity at the gridpoint. Regions with no fields within a distance equal to the weighting $\sigma$ are not given a value; this stops the mapping from extrapolating too far away from the measured fields. 

Figure~\ref{fig:metallicity} shows interpolations across the LMC with Gaussian kernel sizes of $\sigma = 0.15$\,deg and $\sigma = 1$\,deg for the measured gas phase metallicity.  {Two interpolations were chosen: one to show the small-scale structure by focussing on individual fields, the other to map the large-scale structures by combining multiple fields.  The interpolation parameters were chosen to best match the desired scale.}  For the $\sigma = 0.15$\,deg interpolation, the cutoff value means that only a small region around each field is given a value. Although a $\sigma = 0.15$\,deg cut-off is small, some regions have multiple fields close enough that they influence the values displayed. Other regions only have one field within this $\sigma$. The $\sigma = 1$\,deg interpolation expands the regions covered to include areas that have no \Halpha\ emission regions. However, it is easier to appreciate trends in the gas phase metallicity across the LMC in this figure than in the more tightly-focussed $\sigma = 0.15$\,deg interpolation. The map smoothed to 1\,deg is similar to an observation of a high-redshift galaxy, where individual \HII\ regions are not resolved. The area covered by the WiFeS field of view does not always cover the entire \HII\ region being sampled, particularly in the case of 30~Doradus. However, we expect the effect that this sampling has on the metallicity is small, as proximate regions such as in 30~Doradus or N11 show the same metallicities within uncertainties.  Some features in the 1~degree smoothing map may be numerical artefacts due to the uneven spatial sampling of the WiFeS fields used in the interpolation.  However, the large scale of the 1~degree smoothing does ensure, for most of the map, that many WiFeS fields are included in the interpolated values.  The error map in Figure~\ref{fig:metallicity} shows that there is no particular pattern in the metallicity distribution linked to the errors; that is, the fields with the highest errors do not correlate with extremely high or low metallicity values.

The log(O/H)+12 metallicity varies by \around 0.5\,dex across the LMC. This is a relatively large variation and indicates a complex star formation history in the LMC. 30~Doradus is the most active star formation region of the LMC (at RA=84.68\degrees, Dec=$-$69.10\degrees), but does not stand out in the interpolated map. It has a metallicity of 8.25\,dex in the $\sigma = 1$\,deg map. The region with the highest gas phase metallicity lies to the north of the bar, near the region labelled Constellation~III by \citet{harris09}, who find this region only started star formation less than 50~Myr ago. The variation found in LMC gas phase metallicity is large compared to other galaxies, but not unprecedented, as can be seen in the sample of 49 local field star-forming galaxies from \citet{ho15}. We expect stronger variations within smaller galaxies, as the impact of single star formation events can be larger.  Additionally the LMC is interacting strongly with both the Small Magellanic Cloud and the Milky Way which may influence the variations seen.


\begin{table*}
\centering
\caption{The comparison of metallicity values between our work and the literature for fields within the LMC.  We give just the random error on our values and do not include the dispersion due to the method used, which can be up to 0.32~dex according to  \citet{perezmontero09} or 0.25~dex according to \citet{lopezsanchez12}.  {The last column is the difference between the literature value and our log(O/H)+12 value.} }
\label{tab:metallicity_comparison}
\begin{tabular}{cccccccc} 
\hline 
                      &            &      &        & Literature & This Work's & This Work's  & Difference With  \\
Source               & Object     & RA    &  Dec   & log(O/H)+12 & log(O/H)+12 & [NII] & Literature Value\\
\hline
\citet{rolleston02} & LH 9-1160 & 74.14 & -66.48 & 8.29 $\pm$ 0.26 & 8.170 $\pm$ 0.004 & 8.160 $\pm$ 0.004 & 0.120 $\pm$ 0.26 \\
\citet{rolleston02} & LH 10-3270 & 74.34 & -66.42 & 8.28 $\pm$ 0.14 & 8.176 $\pm$ 0.004 & 8.164 $\pm$ 0.005 & 0.104 $\pm$ 0.14 \\
\citet{rolleston02} & LH 104-39 & 85.01 & -69.42 & 8.45 $\pm$ 0.32 & 8.372 $\pm$ 0.027 & 8.385 $\pm$ 0.043 & 0.078 $\pm$ 0.32 \\
\citet{rolleston02} & LH 104-24 & 85.02 & -69.40 & 8.59 $\pm$ 0.19 & 8.362 $\pm$ 0.027 & 8.376 $\pm$ 0.043 & 0.228 $\pm$ 0.19 \\
\cite{toribiosancipriano17} & 30 Doradus & 84.68 & -69.10 & 8.39 $\pm$ 0.01 & 8.194 $\pm$ 0.010 & 8.228 $\pm$ 0.014 & 0.196 $\pm$ 0.014 \\
\cite{toribiosancipriano17} & N44C & 80.56 & -67.98 & 8.31 $\pm$ 0.03 & 8.490 $\pm$ 0.010 & 8.353 $\pm$ 0.012 & 0.180 $\pm$ 0.032 \\
\cite{toribiosancipriano17} & IC 2111 & 72.97 & -69.39 & 8.43 $\pm$ 0.04 & 8.336 $\pm$ 0.008 & 8.313 $\pm$ 0.011 & 0.0944 $\pm$ 0.041 \\
\cite{toribiosancipriano17} & NGC 1714 & 73.04 & -66.92 & 8.37 $\pm$ 0.04 & 8.127 $\pm$ 0.004 & 8.133 $\pm$ 0.004 & 0.243 $\pm$ 0.040 \\
\cite{toribiosancipriano17} & N11B & 74.20 & -66.41 & 8.39 $\pm$ 0.03 & 8.173 $\pm$ 0.004 & 8.160 $\pm$ 0.005 & 0.217 $\pm$ 0.030 \\
\citet{mcleod19} & N44 all & 80.57 & -67.94 & 8.32 $\pm$ 0.11 & 8.490 $\pm$ 0.010 & 8.353 $\pm$ 0.012 & 0.170 $\pm$ 0.11 \\
\citet{crowther23} & 30 Doradus & 84.68 & -69.10 & 7.987 $\pm$ 0.006 & 8.194 $\pm$ 0.010 & 8.228 $\pm$ 0.014 & 0.207 $\pm$ 0.011\\

\end{tabular}
\end{table*}


Comparisons with literature values for individual fields are shown in Table~\ref{tab:metallicity_comparison}.  The comparisons are not from identical locations but for locations in the LMC that are fairly close (within 0.15 degrees), a factor that may partially explain any differences.  The comparison with the OB-type main-sequence stars of \citet{rolleston02} is reasonable considering the size of their errors.  Additionally, the comparison with the empirically derived value for the field of \citet{mcleod19} is within two-sigma of our result.  However, the results of \citet{toribiosancipriano17} from collisionally excited lines differ significantly from our results, considering the errors.  The metallicity values for neighbouring fields in our data are close to one another, so it is unlikely that the difference found with these results is due to a large undetermined random error, and looks instead like a systematic difference.  This may be caused by the dispersion in the empirical method of \citet{pettini04} that we have used, which can be up to 0.32~dex according to \citet{perezmontero09} or 0.25~dex according to \citet{lopezsanchez12}.  The metallicity value listed in the table for \citet{crowther23} is calculated from their integrated line flux that they list in their paper using the same O3N2 conversion used in this work.  This value for 30 Doradus is significantly lower than our value and also significantly lower than that from \citet{toribiosancipriano17}.  It is of interest that our value lie between these two measurements.

No sign of the LMC stellar bar can be seen in the metallicity map, just as it cannot be seen in the distribution of the \HI\ gas \citep{kim98}.  Many authors have attempted to measure a radial gradient in the metallicity of the LMC. \citet{cioni09} studied stellar metallicities of asymptotic giant branch (AGB) stars in the LMC, finding a smooth gradient with a slope of $-0.047 \pm 0.003$\,dex\,kpc$^{-1}$ to \around 8\,kpc.  \citet{choudhury16} confirmed a similar slope from observations of red giant branch (RGB) stars. \citet{feast10} used RR Lyrae variables and found a shallow gradient in stellar metallicity with a slope between $-0.06$ and $-0.04$\,dex (R/R25)$^{-1}$ to beyond a galactocentric radius of 5\,kpc. \citet{pagel78} found a very small negative gradient from \HII\ regions. \citet{toribiosancipriano17} found an essentially flat O/H gradient across the LMC. All these observations have been limited by poor spatial sampling. As can be seen from the gas phase metallicity maps of the LMC, different radial directions have different metallicity gradients, including both negative, flat and positive gradients (the emission line gradients vary as much 0.14\,dex~kpc$^{-1}$). In reality, the spatial distribution of metals in the LMC gas does not follow a simple pattern and cannot be captured by a single radial gradient.  This could be due to the fact that the LMC is an irregular galaxy or due to the fact that it is interacting with both the Small Magellanic Cloud and the Milky Way. 

{The LMC is not a simple irregular galaxy, as it shows evidence for being a dwarf spiral such as a distinct stellar bar and spiral arms \citep{geyer77,cioni00,wilcots09}.  It has been suggested that the LMC was once a low surface brightness galaxy that, due to interactions with the Small Magellanic Cloud, evolved into the more irregular shape it now has \citep{vandenbergh97}.  This complex history may cause the complex structure seen in the metallicity maps.}

We also determined the gas phase metallicity using the N2 calibration calculated for the LMC fields using the formula from \citet{pettini04}:
\begin{equation}
\rm 12 + \log(O/H) = 9.37 + 2.03N2 + 1.26N2^2 + 0.32N2^3
\end{equation}
where $\rm N2 = log([NII]\lambda 6584/H\alpha)$.  A conversion from \citet{kewley08} was used to calibrate these estimates to the above O3N2 estimates and allow direct comparisons:
\begin{equation}
\rm 12 + \log(O/H) = -8.0069 + 2.74353 x + -0.093680 x^2
\end{equation}
where $x$ is the original N2 metallicity measurement. The N2 metallicity for the fields ranges from 8.08\,dex to 8.44\,dex and its mean is 8.30\,dex.  Figure~\ref{fig:metallicity_NII} shows smoothed maps created the same way as the above maps using the O3N2 calibration.  In most locations the maps are in good agreement, except in the northern region that shows higher metallicities in the O3N2 map.  This feature, while present in the N2 map, does not have as high a metallicity.  A possible origin of this difference is that N2 has a lower dependence on the ionisation parameter \citep{kewley19}.

{The fluxes of the [S{\sc ii}] emission lines were measured for the fields and the ratio compared to the electron density relationship.  The line ratios for our data all lay around 1.4, which is in the regime where the line ratio to electron density relationship is fairly flat.  This regime runs from electron density 1 to 100 cm$^{-3}$, and so all we can say is that the electron density is fairly low.}     

\subsection{Extinction}
\label{Extinction}

An even smaller sample of 28 fields were selected for extinction measurements. Extinction was measured by looking at the ratio of the fluxes of the \Halpha\ and \Hbeta\ emission lines. To be selected in this sample, each \Halpha\ and \Hbeta\ emission line measurement had to have a Gaussian FWHM $>0.5$\AA\ and $<2$\AA\ and a signal-to-noise in the line~$\ge 3$. As we are now directly comparing a value in the blue spectrum (\Hbeta) with a value in the red spectrum (\Halpha), we have to worry about having a consistent flux calibration in both spectra. Unfortunately, the flux calibration was not always successful, and there are values for the ratio of \Halpha\ and \Hbeta\ that are not physically possible. Anomalous values were removed based on their calculated extinction values, as described below. To calculate the extinctions (measured in magnitudes) we use
\begin{equation}
\rm E(B-V)_{H\beta - H\alpha} = \frac{-2.5}{k_{H\beta} - k_{H\alpha}} log \frac{(H\alpha/H\beta)_{int}}{(H\alpha/H\beta)_{obs}}
\end{equation}
where $\rm (H\alpha/H\beta)_{int}$ is the intrinsic flux of ratio of \Hbeta\ to \Halpha, taken to be 2.86 assuming case~B recombination with a temperature of 10,000\,K and an electron density of 100\,cm$^{-3}$ \citep{osterbrock06}.  $\rm (H\alpha/H\beta)_{obs}$ is the observed flux ratio of \Hbeta\ to \Halpha. k$\rm _{H\beta}$ and k$\rm _{H\alpha}$ are taken from the average LMC extinction curve of \citet{gordon03}. The exact values at the wavelength of \Hbeta\ and \Halpha\ are taken from a spline fit to the \citet{gordon03} data and are k$\rm _{H\beta}$ = 3.97 and k$\rm _{H\alpha}$ = 2.56. 

To be accepted, a measurement had to have $\rm E(B-V)_{H\beta - H\alpha} > 0$ and $\rm E(B-V)_{H\beta - H\alpha} < 0.4$ and an error in $\rm E(B-V)_{H\beta - H\alpha}$ that is $<0.1$.  Before making these cuts there were 64 fields, which dropped to 42 when applying the error cut.  There were 2 fields with anomalously high values that were easy to identify, as they were they both had an $\rm E(B-V)_{H\beta - H\alpha}$ higher than 0.8, well above the expected limit of 0.4 (see \citet{chen22}).  In addition, there were 12 negative $\rm E(B-V)_{H\beta - H\alpha}$ values, which also are anomalous. In these cases it was easy to identify that the flux calibration between the blue and red spectrum was causing problems.  After these cuts there were 28 good fields left.  While this is a small number, the fields were well spread across the LMC, allowing for a reasonable sampling.  The extinction values for these fields are given in Table~\ref{tab:EBV}. It should be noted that uncertainties in the flux calibration between the blue and red spectrum may not always lead to a dramatic difference, so there may be further errors in the \Hbeta\ to \Halpha\ flux ratio that could affect these results.

\begin{figure}
  \includegraphics[width=\columnwidth]{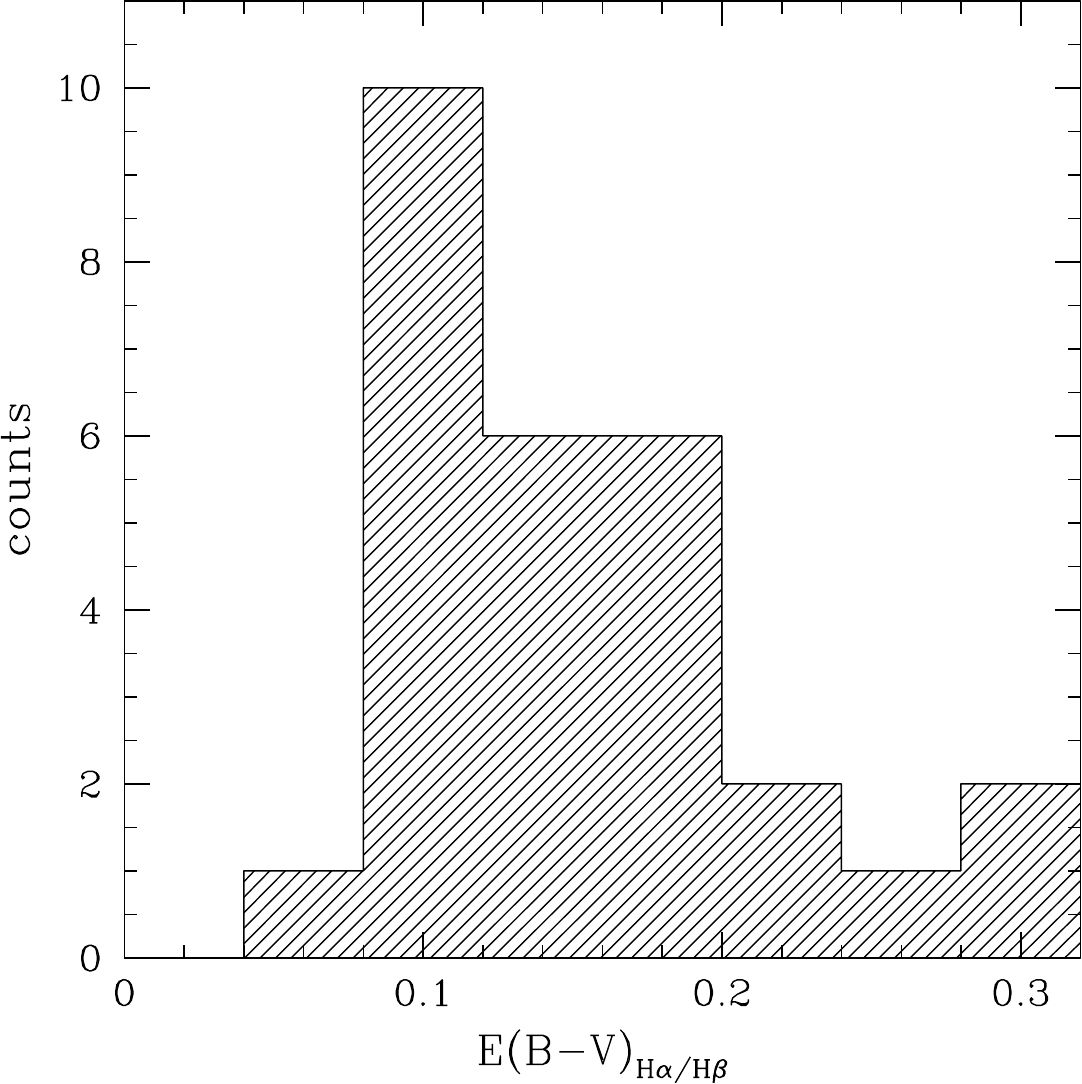}
  \caption{Histogram of $\rm E(B-V)_{H\beta - H\alpha}$ for the 28 WiFeS fields.}
  \label{fig:hist_emission_lines_EBV_Halpha_Hbeta}
\end{figure}

Figure~\ref{fig:hist_emission_lines_EBV_Halpha_Hbeta} is a histogram of $\rm E(B-V)_{H\beta - H\alpha}$ for each of these 28 WiFeS fields. \citet{gorski20} made an extinction map across the LMC  based on red clump stars selected from the OGLE-III photometric data. The mean value of the reddening they found for the LMC was E(B-V)~=~$0.127 \pm 0.013$\,mag. The mean for our fields is $E(B-V)_{H\beta - H\alpha}$~=~0.157\,mag, which is not too dissimilar.  \citet{cox06} measured the extinction in the LMC in 12 diffuse interstellar bands in five lines of sights to early-type stars in the LMC, and found E(B-V) in the range between 0.1 to 0.4.


\begin{figure*}
  \includegraphics[width=\columnwidth]{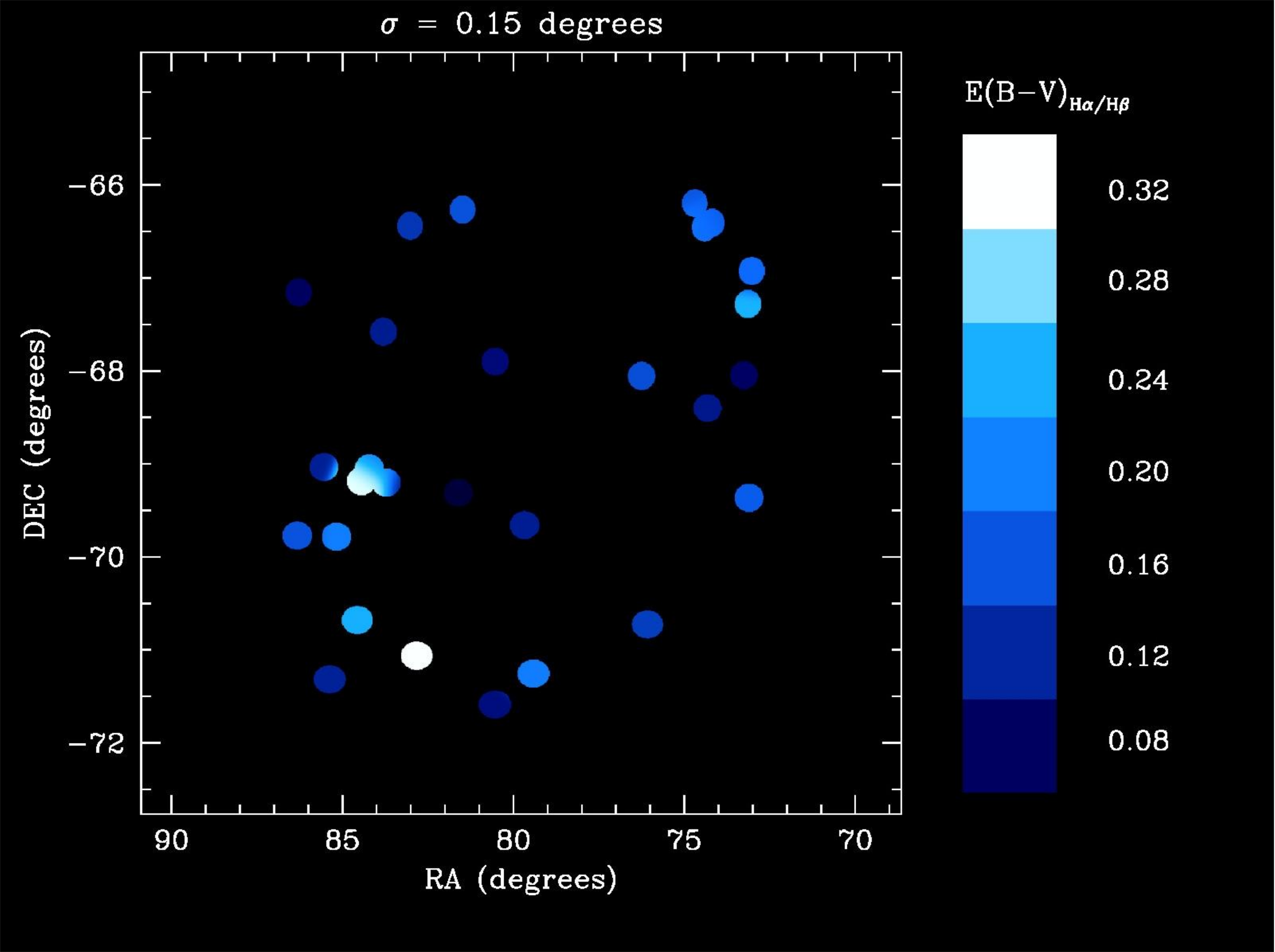}
  \includegraphics[width=\columnwidth]{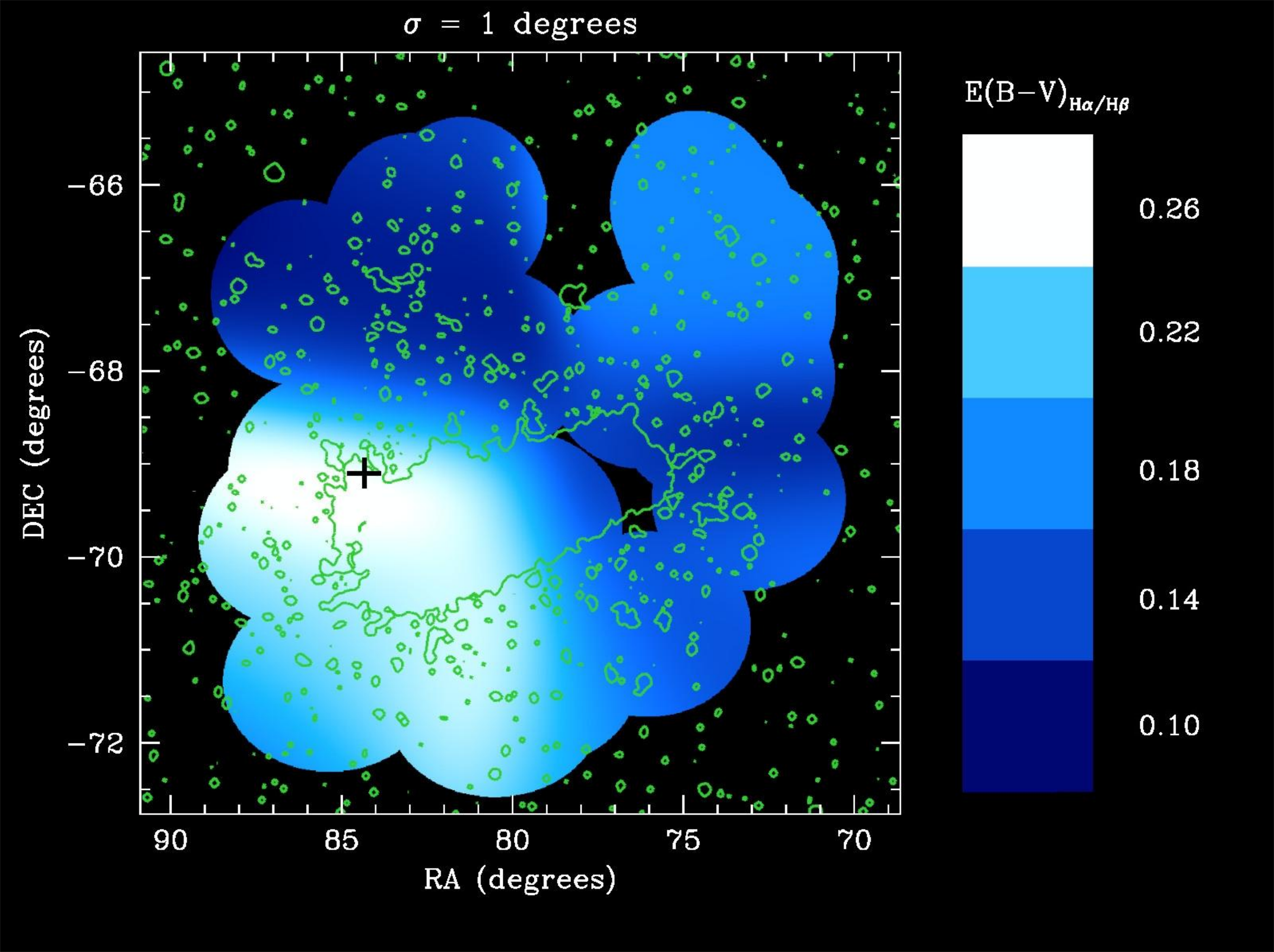}
  \caption{Maps of $\rm E(B-V)_{H\beta - H\alpha}$ across the LMC, interpolated using a Gaussian kernel with $\sigma = 0.15$\,deg in the left panel and $\sigma = 1$\,deg in the right panel; the interpolation has a cutoff distance from the nearest field equal to the $\sigma$ value. The contours in the right panel are from the SHASSA Continuum Image and highlight the LMC stellar bar.  The black cross in the right panel is the location of 30 Doradus.}
  \label{fig:EBV}
\end{figure*}

Interpolation maps for extinction, like those for metallicity, are shown in Figure~\ref{fig:EBV}, with the left panel having a resolution of $\sigma = 0.15$\,deg and the right panel having resolution $\sigma = 1$\,deg. There is an extinction peak near the location of 30~Doradus (at RA=84.68\degrees, Dec=-69.10\degrees).  \citet{cox06}, \citet{imara07}, \citet{haschke11} and \citet{joshi19} also note that 30~Doradus is a region of high extinction in their reddening maps. In our map 30~Doradus has $\rm E(B-V)_{H\beta - H\alpha} = 0.3$, which is lower than the E(B-V)~=~0.4 found by \citet{cox06} using diffuse interstellar bands in the lines of sight towards early-type stars. The difference in methods for determining the extinction, and the fact that we are looking at $\rm E(B-V)_{H\beta - H\alpha}$ rather than E(B-V), may explain the slightly different estimates. 


We attempted to make direct comparisons of the extinction for individual fields with the stellar maps of \citet{chen22} and \citet{skowron21}, but the errors for these literature samples on the small scales that we were probing was so large that they readily agreed with the extinction we measured.  


\subsection{Radial Velocity}
\label{Radial_Velocity}


\begin{figure}
  \includegraphics[width=\columnwidth]{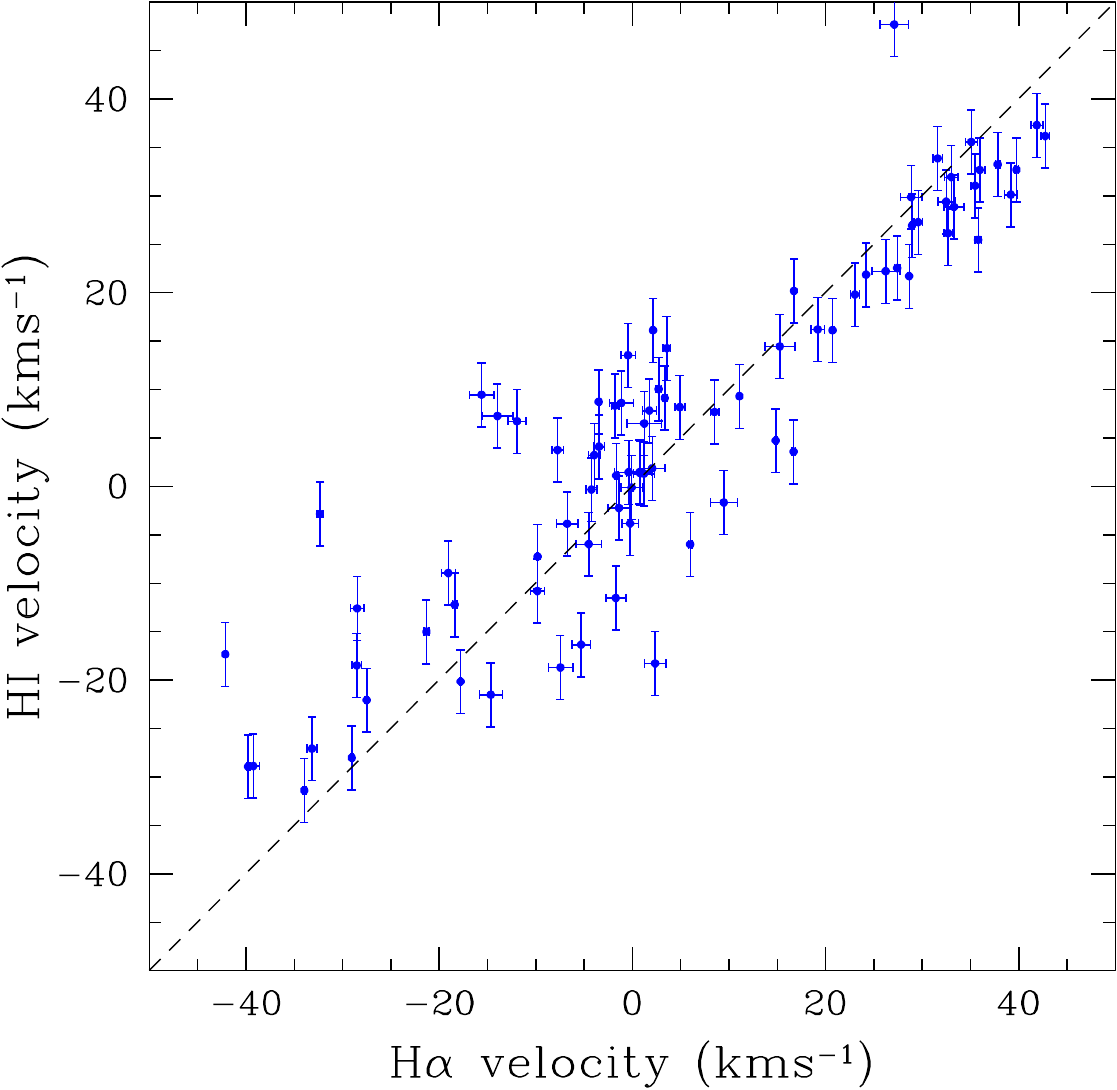}
  \caption{Comparison of the \Halpha\ and \HI\ radial velocities at each of the 83 WiFeS fields.  The systemic velocity for the LMC of 270\kms\ has been subtracted from the radial velocities.}
  \label{fig:plot_compare_vel}
\end{figure}


\begin{figure}
  \includegraphics[width=0.99\columnwidth]{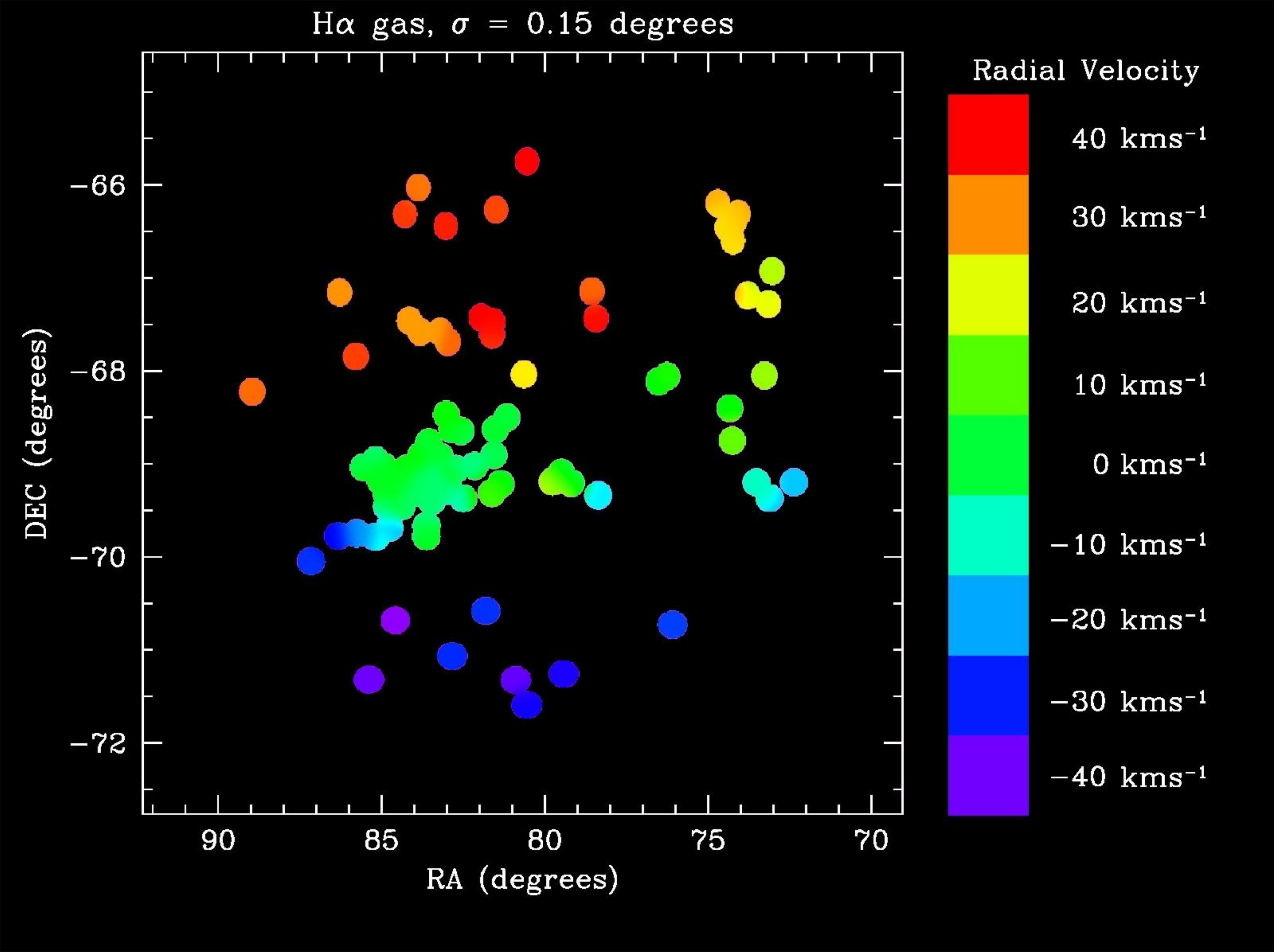}\\
  \includegraphics[width=0.99\columnwidth]{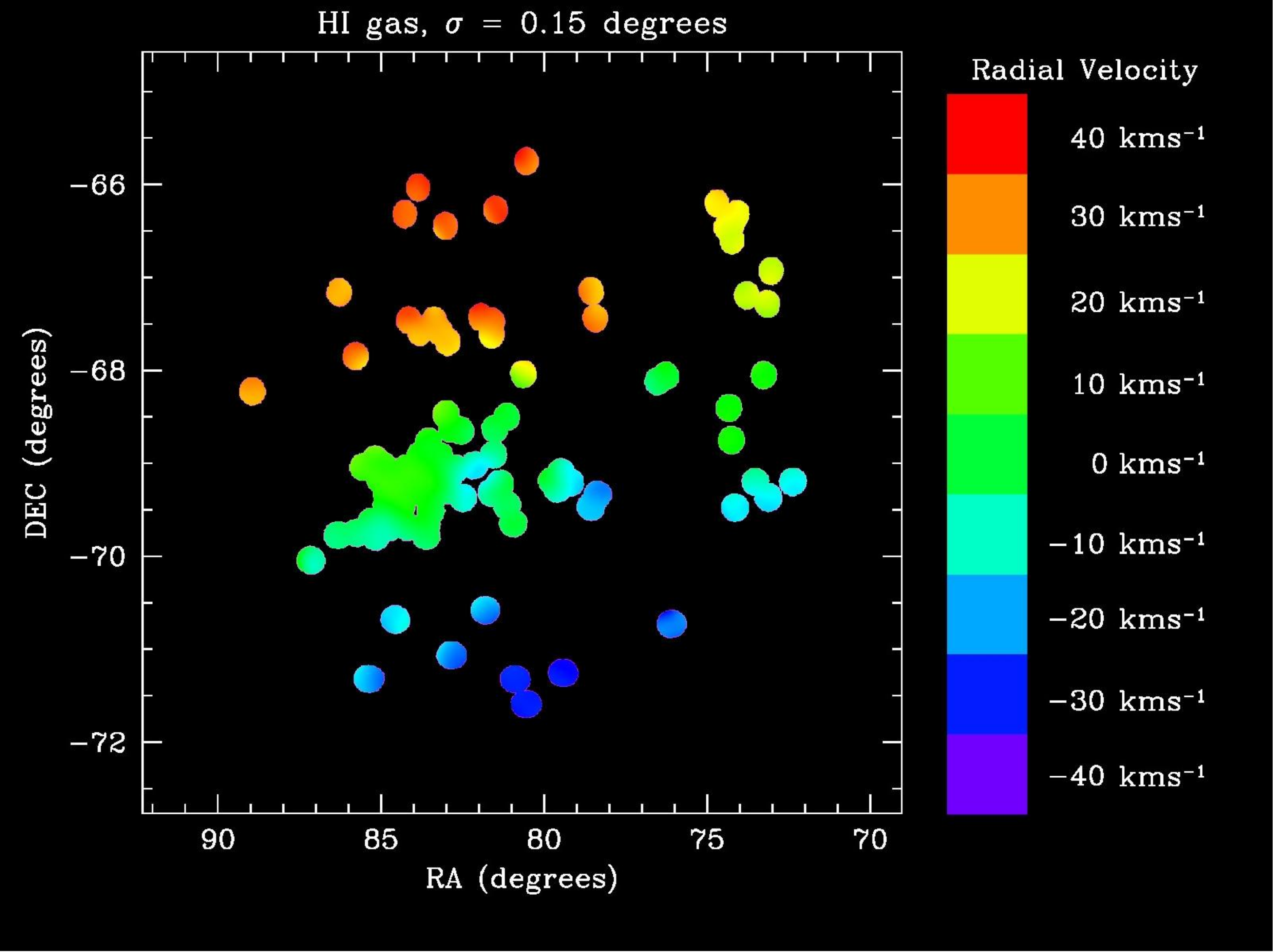}\\
  \includegraphics[width=0.99\columnwidth]{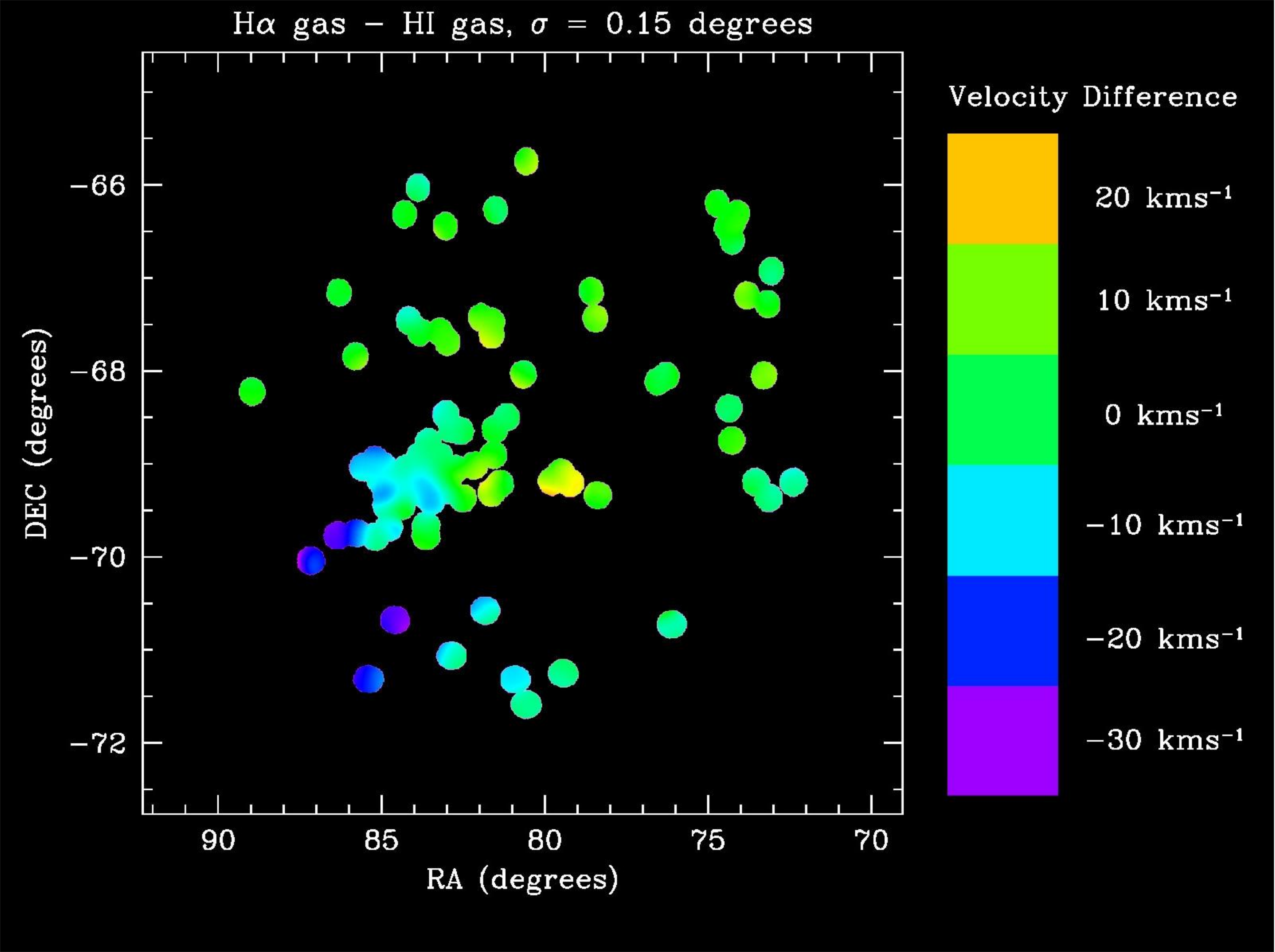}
  \caption{The top panel is the radial velocity across the LMC derived from the \Halpha\ emission line, interpolated using a Gaussian kernel with $\sigma = 0.15$\,deg; the interpolation has a cutoff distance from the nearest field equal to the $\sigma$ value. The middle panel is similar, but shows the \HI\ radial velocity smoothed to $\sigma = 0.15$\,deg over the same region as the top panel. The bottom panel shows the residuals if the \HI\ radial velocity map is subtracted from the \Halpha\ map. The figures are resampled so they are linear in right ascension and declination.}
  \label{fig:radial_velocity_focussed}
\end{figure}


\begin{figure}
  \includegraphics[width=0.99\columnwidth]{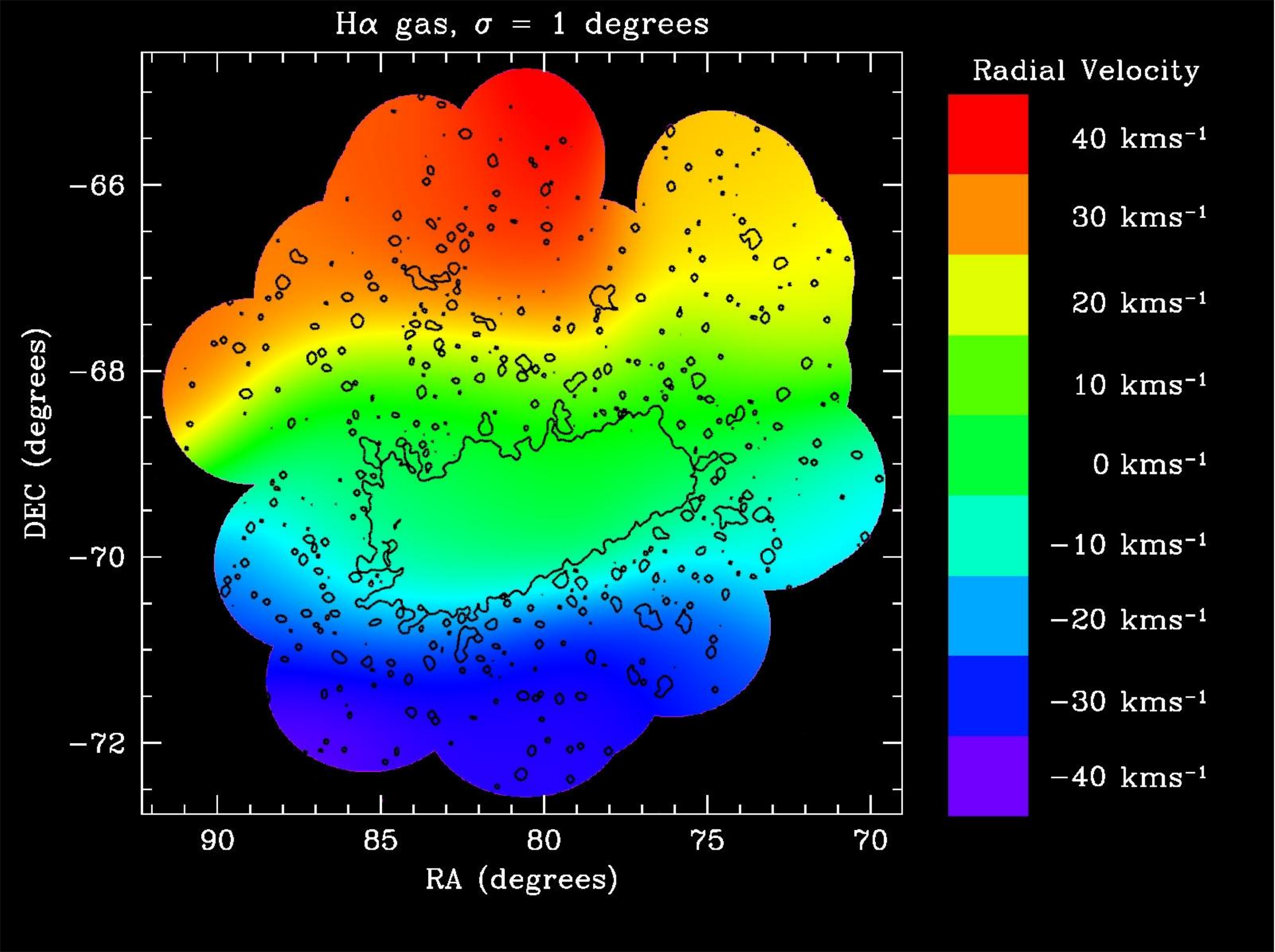}\\
  \includegraphics[width=0.99\columnwidth]{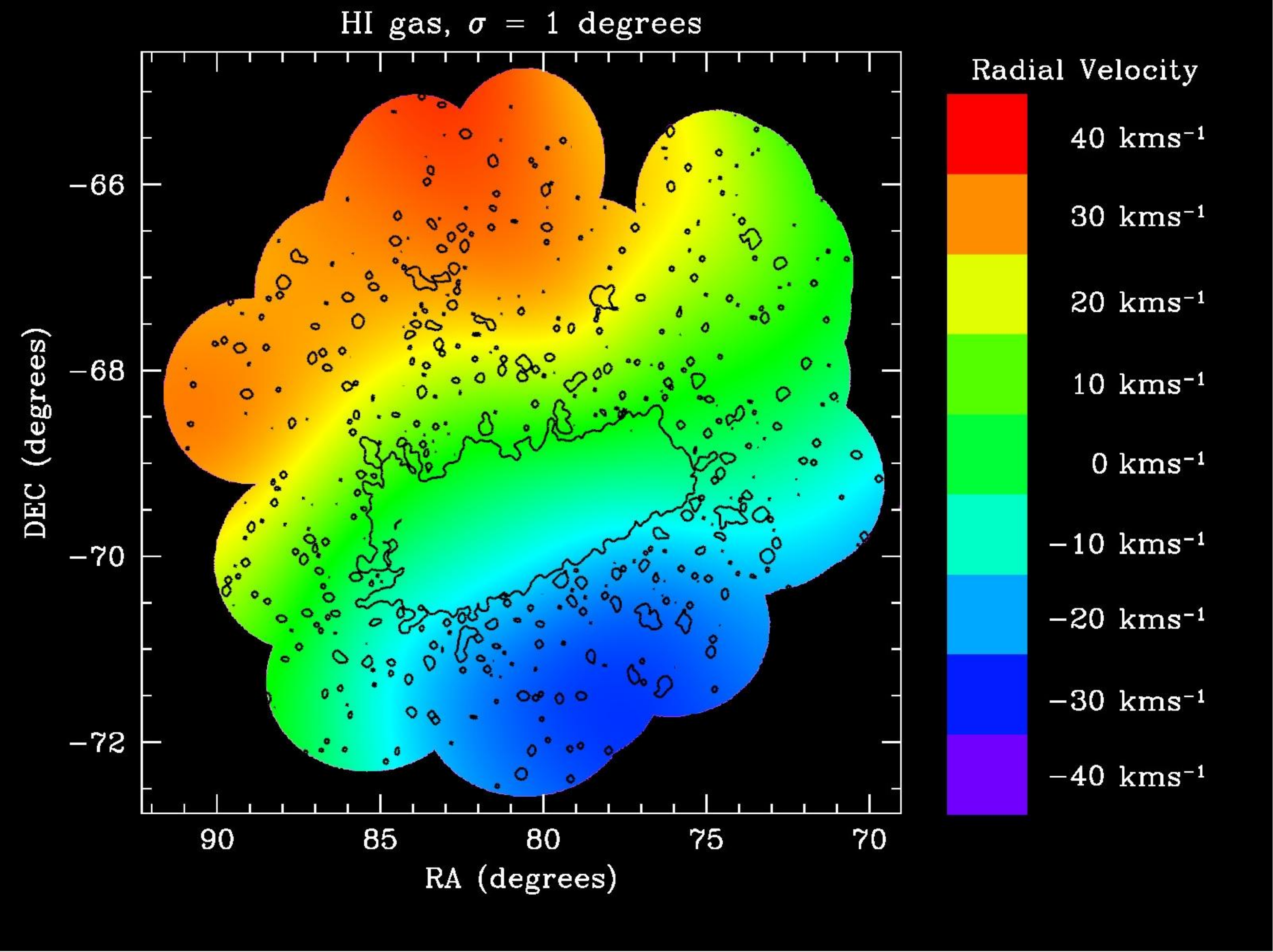}\\
  \includegraphics[width=0.99\columnwidth]{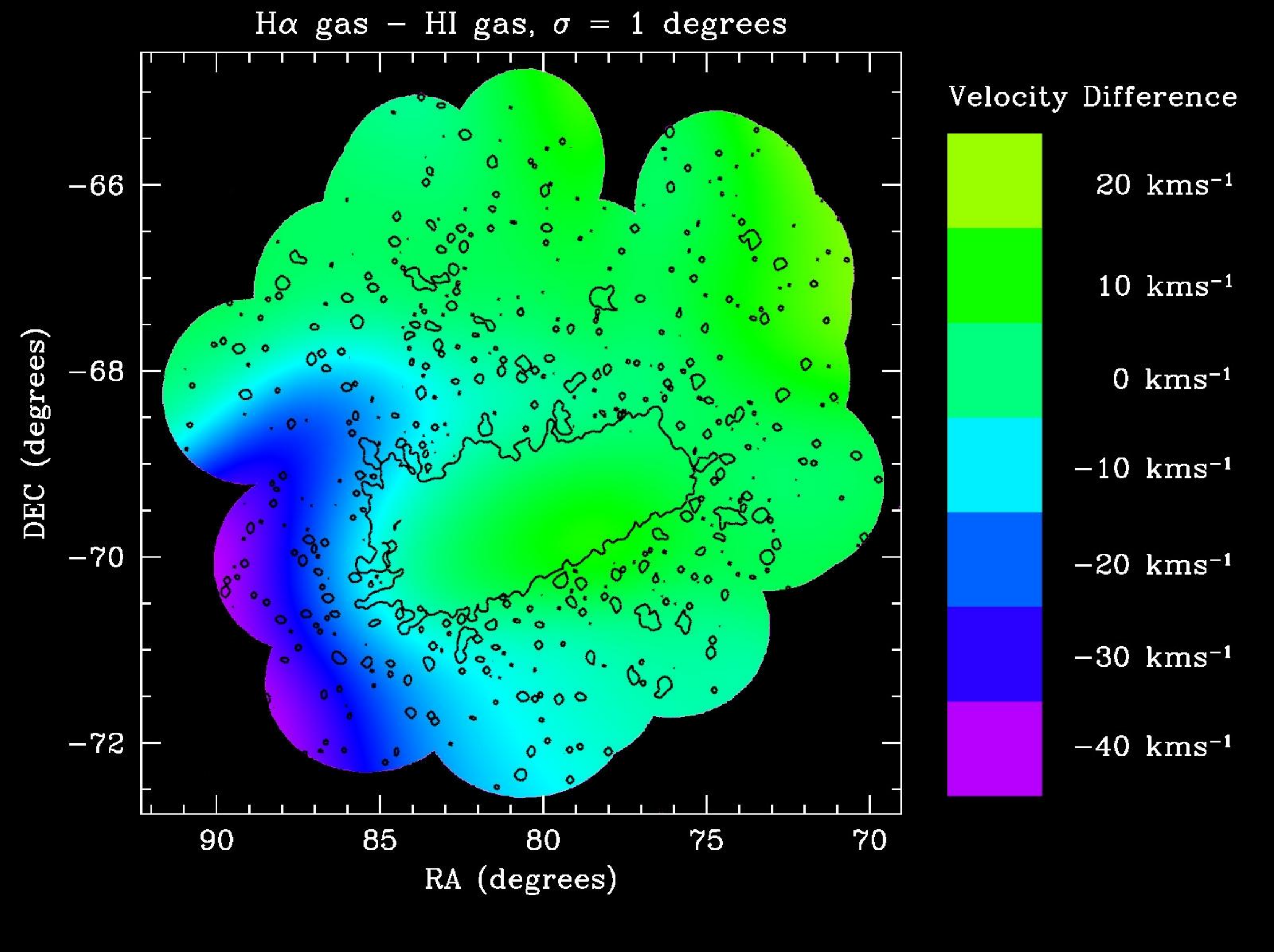}
  \caption{The top panel is the radial velocity across the LMC derived from the \Halpha\ emission line, interpolated using a Gaussian kernel with $\sigma = 1$\,deg; the interpolation has a cutoff distance from the nearest field equal to the $\sigma$ value. The middle panel is similar, but shows the \HI\ radial velocity smoothed to $\sigma = 1$\,deg over the same region as the top panel. The bottom panel shows the residuals if the \HI\ radial velocity map is subtracted from the \Halpha\ map. The figures are resampled so they are linear in right ascension and declination. The contours in all panels are from the SHASSA continuum image and highlight the location of the LMC stellar bar.}
  \label{fig:radial_velocity}
\end{figure}


The radial velocity for each of the observed fields was determined from the \Halpha\ line. Measurements of the sky lines at 6300.304\AA\ and 6863.955\AA\ were used to correct an offset found in the measured \Halpha\ line. After the correction, a small systematic offset ranging from 0 to 2.5\kms\ in the position of the \Halpha\ line was found, but this was deemed sufficiently accurate, given the pixel size was 20\kms\ (the effective velocity resolution is 42.9\kms). The \Halpha\ radial velocity is compared to the radial velocity measured from \HI\ at each of the WiFeS fields in Figure~\ref{fig:plot_compare_vel}. The two radial velocities are generally close, indicating they track the same overall rotating velocity field. The remaining differences are more easily explored in two dimensions, {where they can be localised to particular regions of the LMC}.

Interpolation maps similar to those for gas phase metallicity were made of the \Halpha\ radial velocity, with the top panel of Figure~\ref{fig:radial_velocity_focussed} having $\sigma = 0.15$\,deg and that of Figure~\ref{fig:radial_velocity} having $\sigma = 1$\,deg. A radial velocity of 270\kms\ was adopted as the systemic velocity of the LMC and subtracted from the measured radial velocity. {This value was determined from the data as approximately the midpoint of the velocty distribution.} The \Halpha\ radial velocities are blue-shifted in the region below the LMC bar and red-shifted above and to the north-east of the bar, broadly consistent with rotation around the long axis of the bar.

\begin{figure}
  \includegraphics[width=0.99\columnwidth]{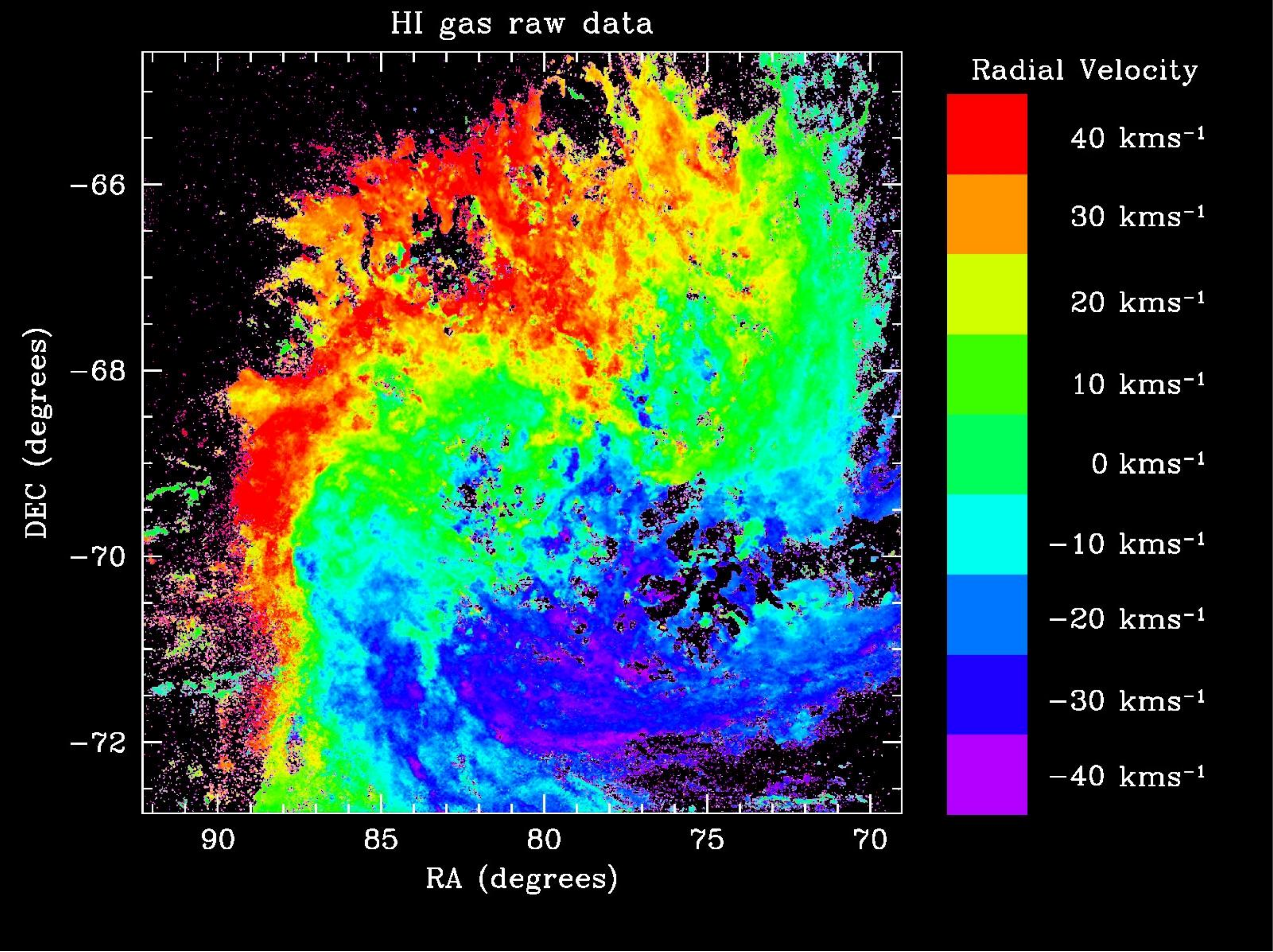}\\
  \includegraphics[width=0.99\columnwidth]{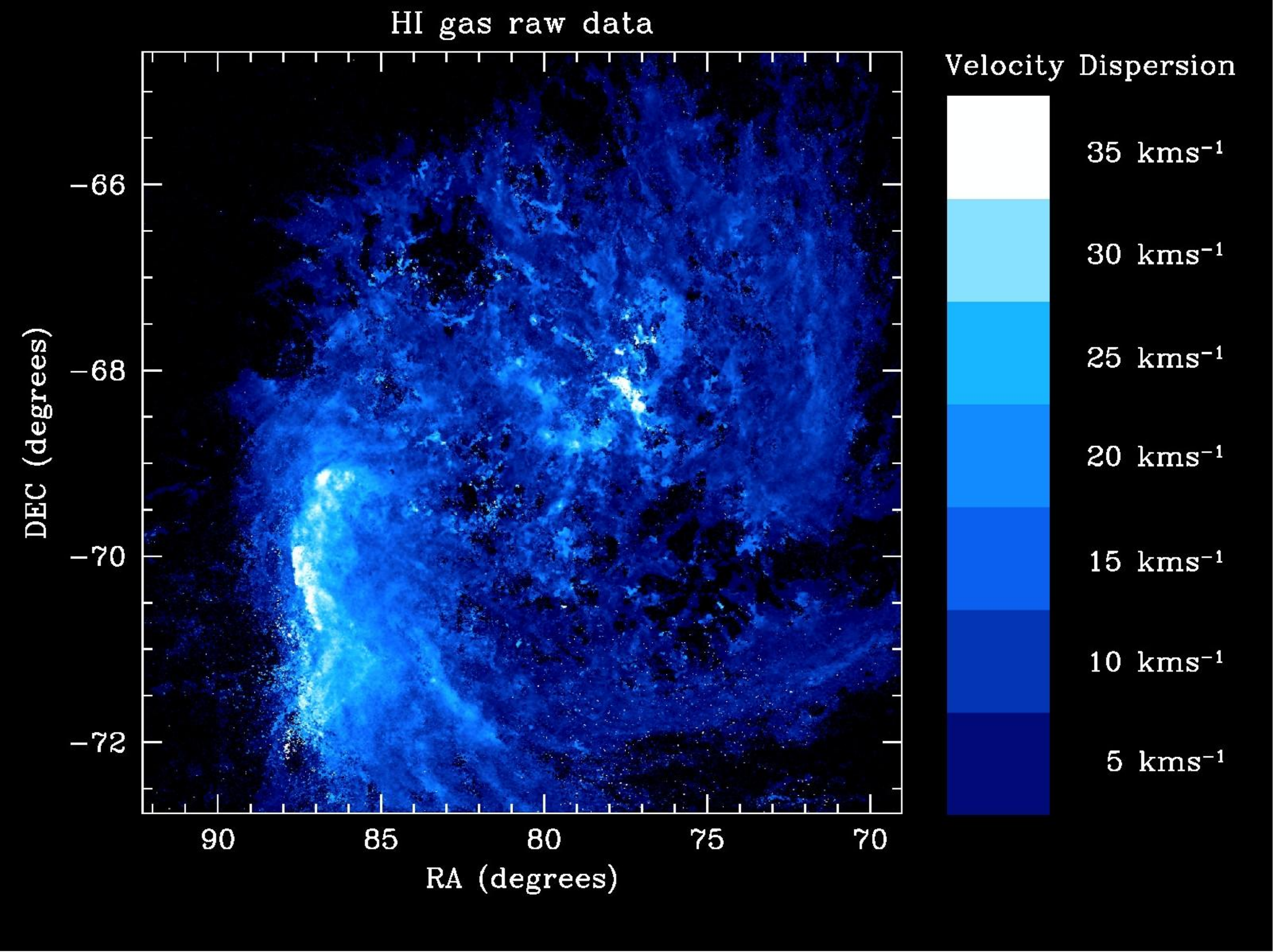}
  \caption{The raw \HI\ radial velocity map (top) and \HI\ velocity dispersion map (bottom) as observed by \citet{kim03} and \citet{staveleysmith03}. The spatial resolution is 1~arcmin.}
  \label{fig:HI_raw}
\end{figure}

The \HI\ radial velocities are shown in the middle panels of Figure~\ref{fig:radial_velocity_focussed} and Figure~\ref{fig:radial_velocity}, smoothed to the same resolution as the equivalent \Halpha\ radial velocity maps and sampling the same regions as the WiFeS observations. The \HI\ data comes from combined Australian Telescope Compact Array (ATCA) and Parkes observations from \citet{kim03} and \citet{staveleysmith03}. The spatial resolution of the combined data set is 1~arcmin and its velocity resolution is 1.65\kms. The raw \HI\ radial velocity data (moment 1 map) at this resolution is shown for reference in the top panel of Figure~\ref{fig:HI_raw}, covering the same limits as our observations. 

The difference between the \Halpha\ and \HI\ radial velocities is shown in the bottom panels of Figure~\ref{fig:radial_velocity_focussed} and Figure~\ref{fig:radial_velocity}. For most of the covered area of the LMC, the \Halpha\ and \HI\ radial velocities are similar. The exception is a region to the south-east, where the \Halpha\ radial velocities are significantly lower (by more than 25\kms)  This could be the \Halpha\ emitting gas being blue-shifted from the \HI\ gas or the \HI\ gas being red-shifted from the \Halpha\ emitting gas. 30~Doradus is just on the boundary of this region. We discuss this region in more detail after considering the velocity dispersion in the next section. 


\subsection{Velocity Dispersion}
\label{Velocity_Dispersion}


\begin{figure}
  \includegraphics[width=\columnwidth]{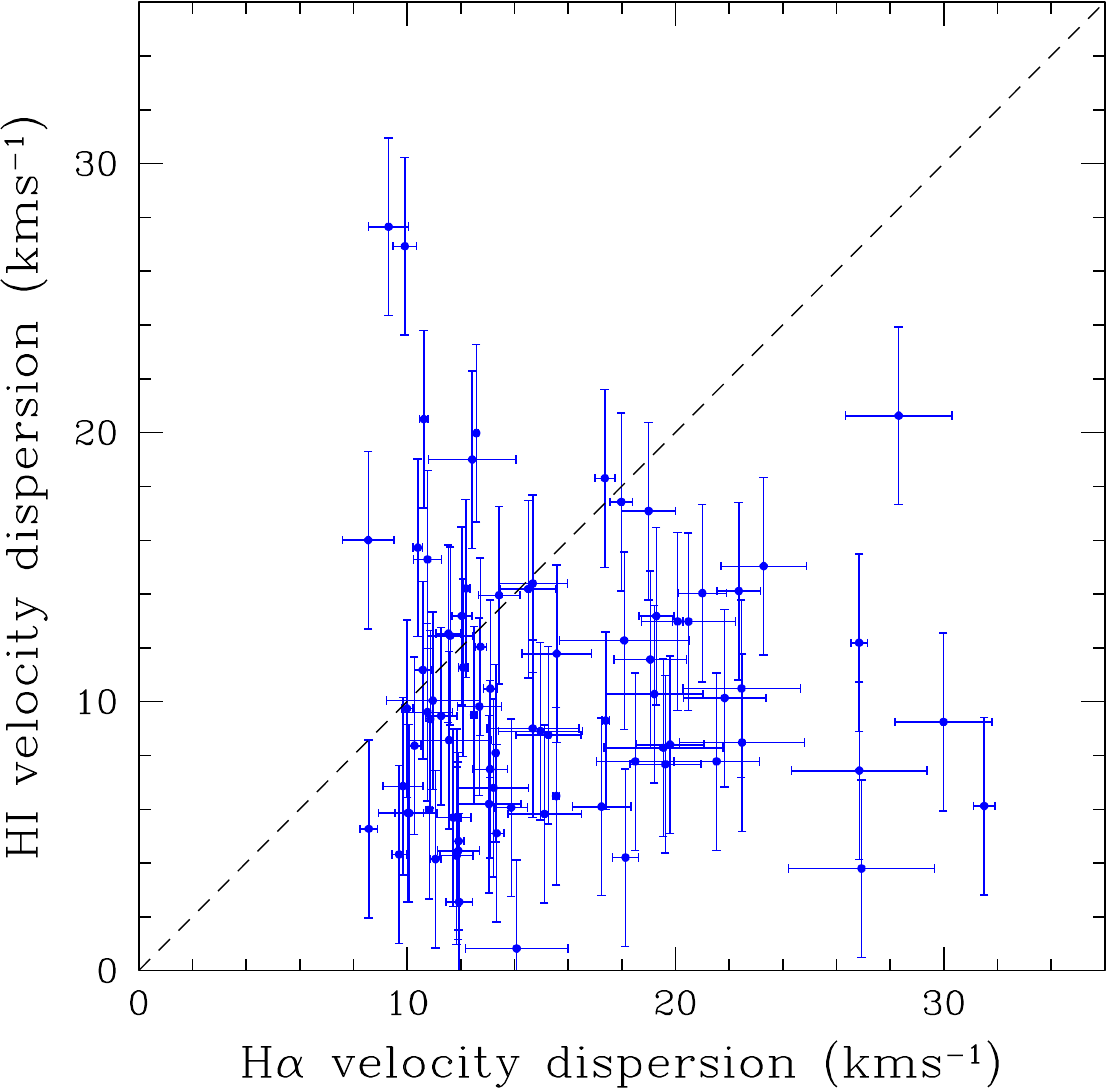}
  \caption{Comparison of velocity dispersion derived from the \Halpha\ and \HI\ spectral lines in each of the 83 WiFeS fields.}
  \label{fig:plot_compare_vd}
\end{figure}

The velocity dispersion has been measured from the width of the \Halpha\ emission line for each of the fields using
\begin{equation}
\rm \sigma_{vd} = \frac{ c\sqrt{\sigma_{H\alpha}^2 - \sigma_{sky}^2} }{ 2.355\lambda(1+z)} 
\end{equation}
where $c$ is the speed of light, $\lambda$ the wavelength of \Halpha\ (6562.8\AA), $z$ the redshift of the emission line, $\rm \sigma_{H\alpha}$ the Gaussian full width half maximum (FWHM) of the \Halpha\ emission line, and $\rm \sigma_{sky}$ the estimated instrumental resolution of WiFeS at the wavelength of \Halpha\ (derived by linear interpolation of the FWHM of the strong sky lines at 6300.304\AA\ and 6863.955\AA).

The velocity dispersion derived from the \Halpha\ line at each WiFeS field is compared to the \HI\ velocity dispersion from \citet{kim03} for each field in Figure~\ref{fig:plot_compare_vd}. As can be seen, the velocity dispersion from \Halpha\ is generally larger than that for \HI, as expected, since the \Halpha\ emitting gas is hot while the \HI\ gas is relatively cold. {The lack of correlation between the velocity dispersions from \HI\ and \Halpha\ is another indication that the processes affecting each gas phase are different.} Our measured velocity dispersions vary across the LMC between 8.5 and 31.5\kms. The larger variation seen in \citet{ambrociocruz16} was due to supernova remnants and superbubbles undergoing expansion motions. We do not sample any of these regions, so do not find their higher values for velocity dispersions.

\begin{figure}
  \includegraphics[width=0.99\columnwidth]{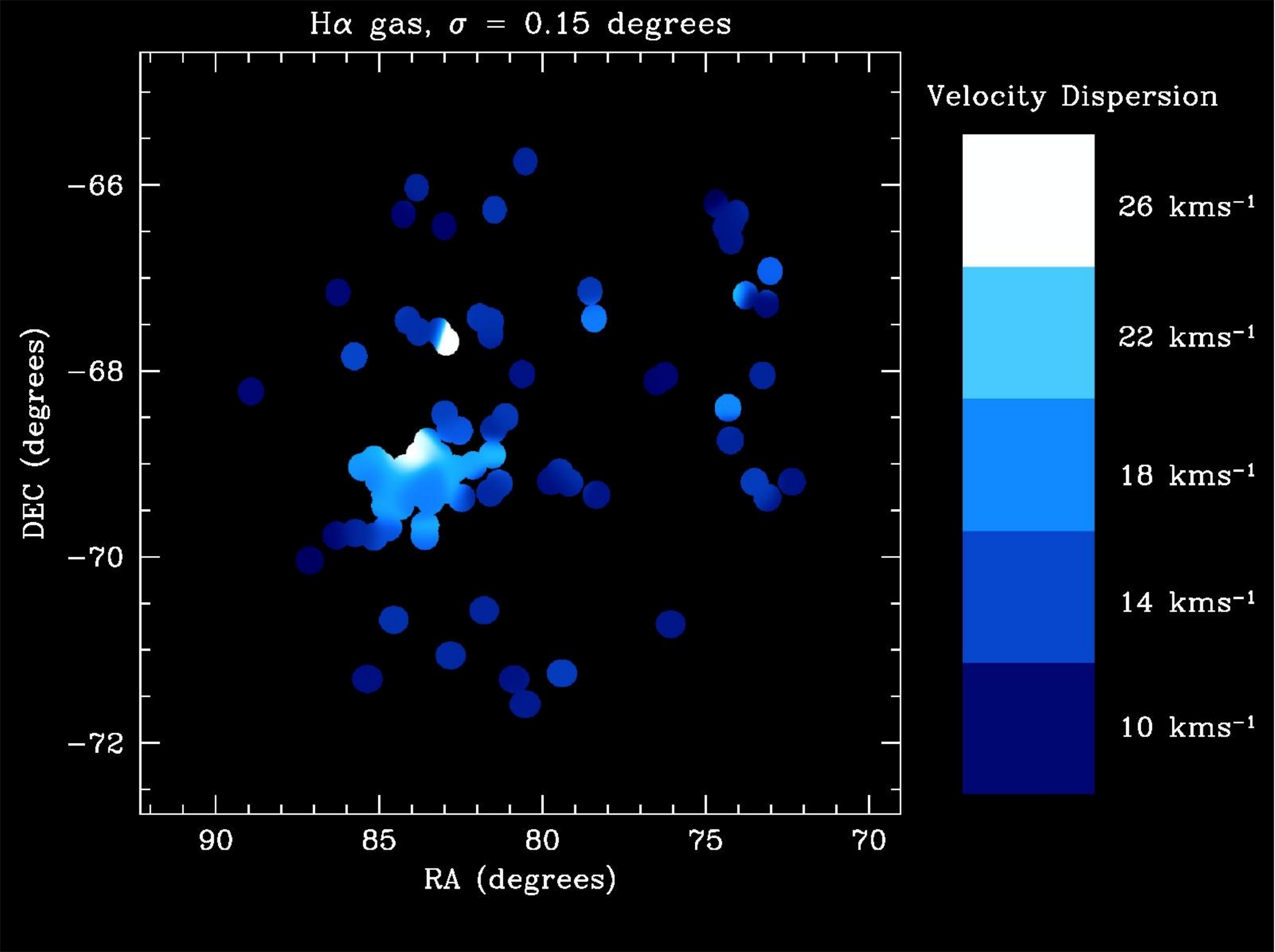}\\
  \includegraphics[width=0.99\columnwidth]{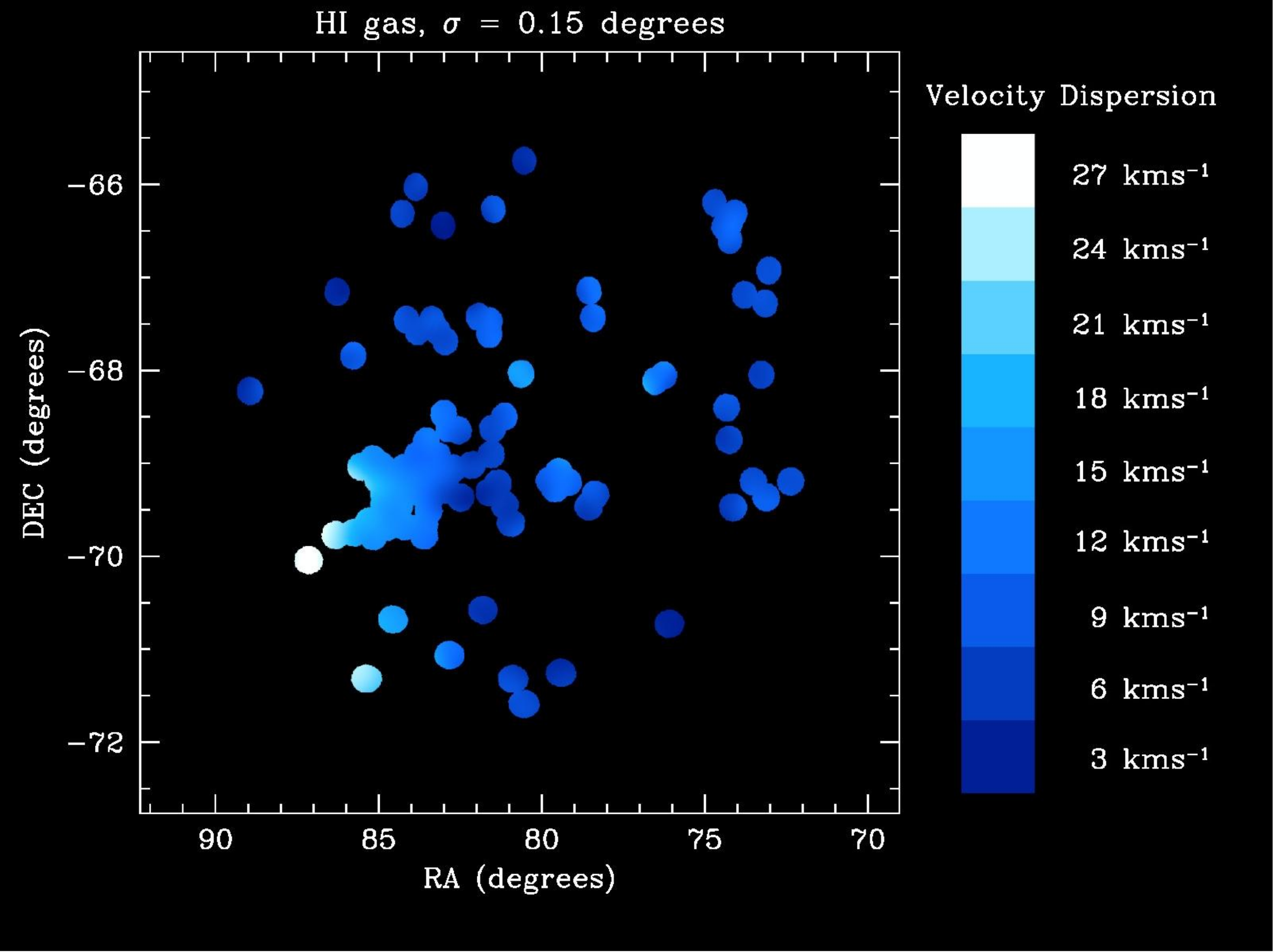}\\
  \includegraphics[width=0.99\columnwidth]{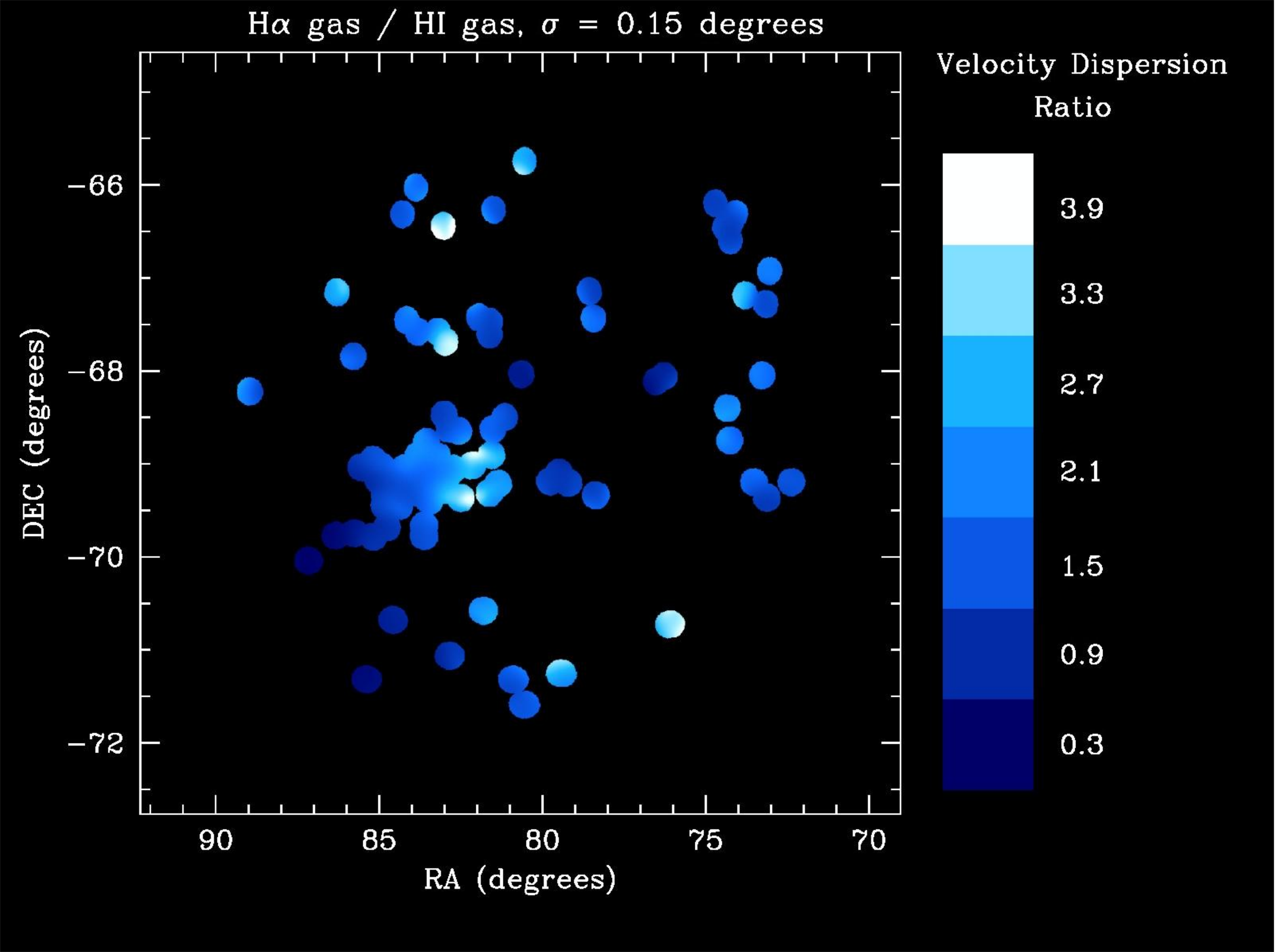}
  \caption{The top panel is the interpolation across the LMC of the velocity dispersion derived from the \Halpha\ emission line using a Gaussian of $\sigma = 0.15$\,deg; the interpolation has a cutoff distance from the nearest field equal to the $\sigma$ value. The middle panel is similar, except it shows the \HI\ velocity dispersion smoothed to $\sigma = 0.15$\,deg over the same region as the top panel. The bottom panel shows the ratio of the \Halpha\ velocity dispersion map divided by \HI\ map. The figures are resampled so they are linear in right ascension and declination.}
  \label{fig:velocity_dispersion_focussed}
\end{figure}

\begin{figure}
  \includegraphics[width=0.99\columnwidth]{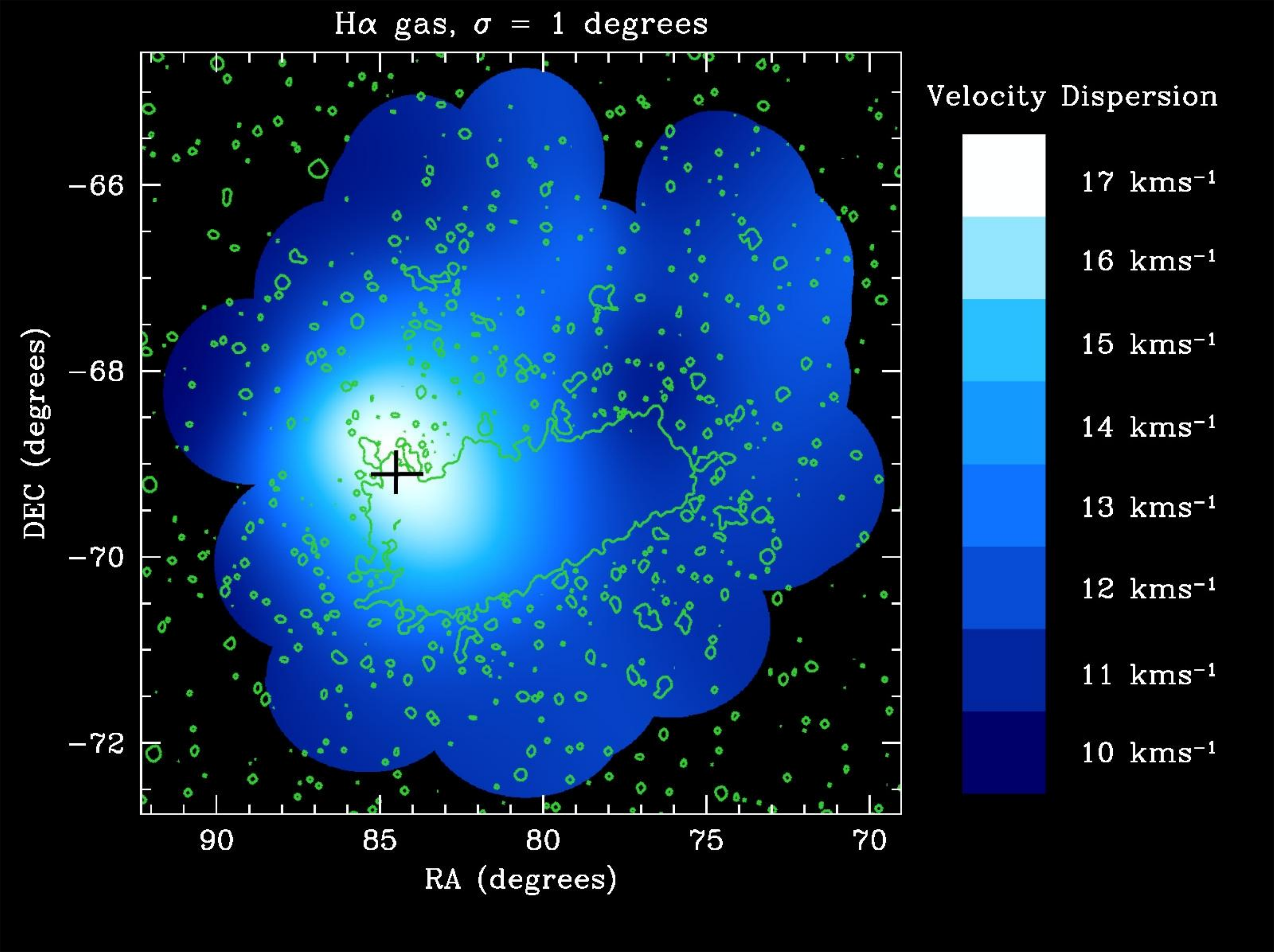}\\
  \includegraphics[width=0.99\columnwidth]{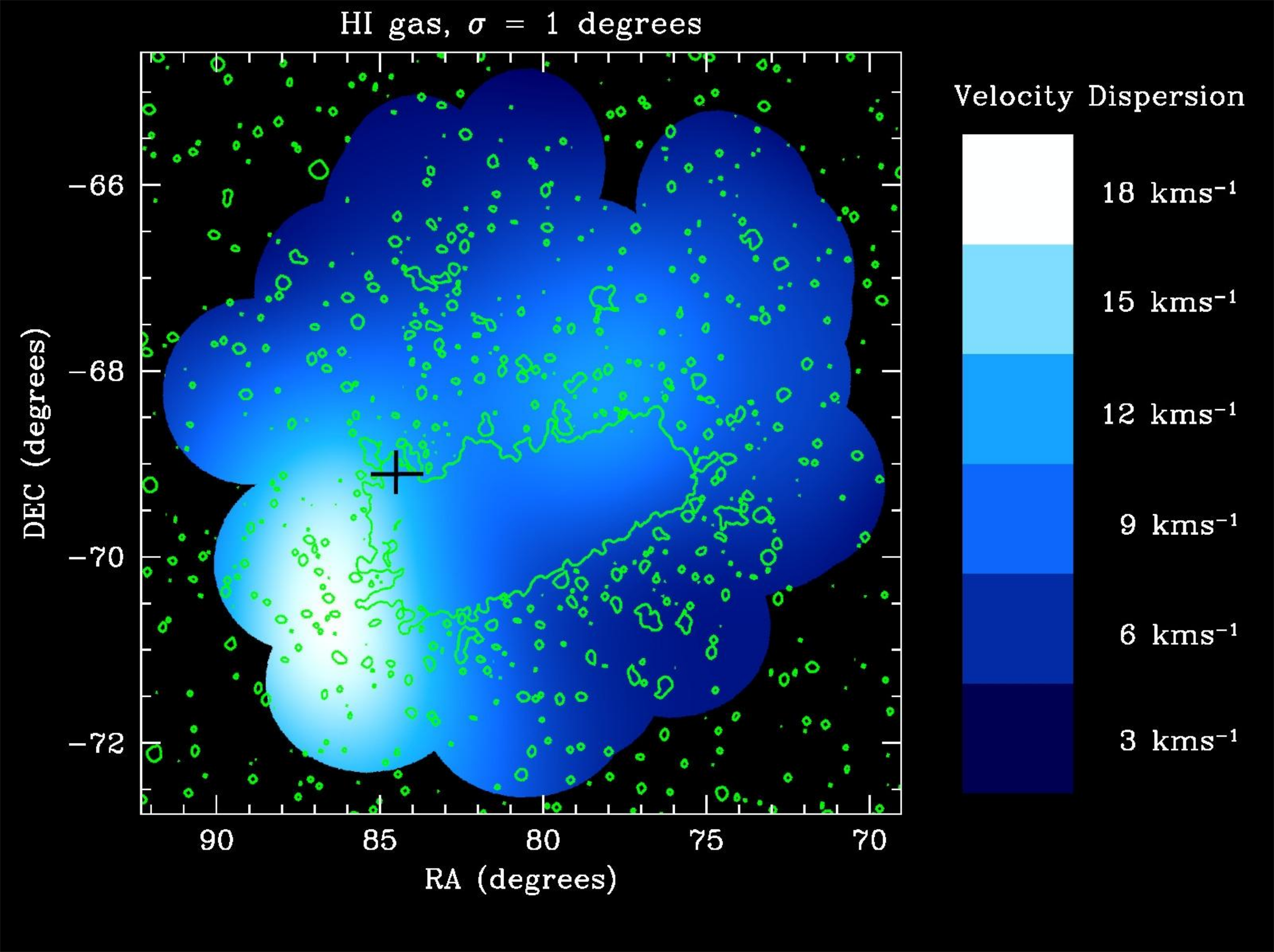}\\
  \includegraphics[width=0.99\columnwidth]{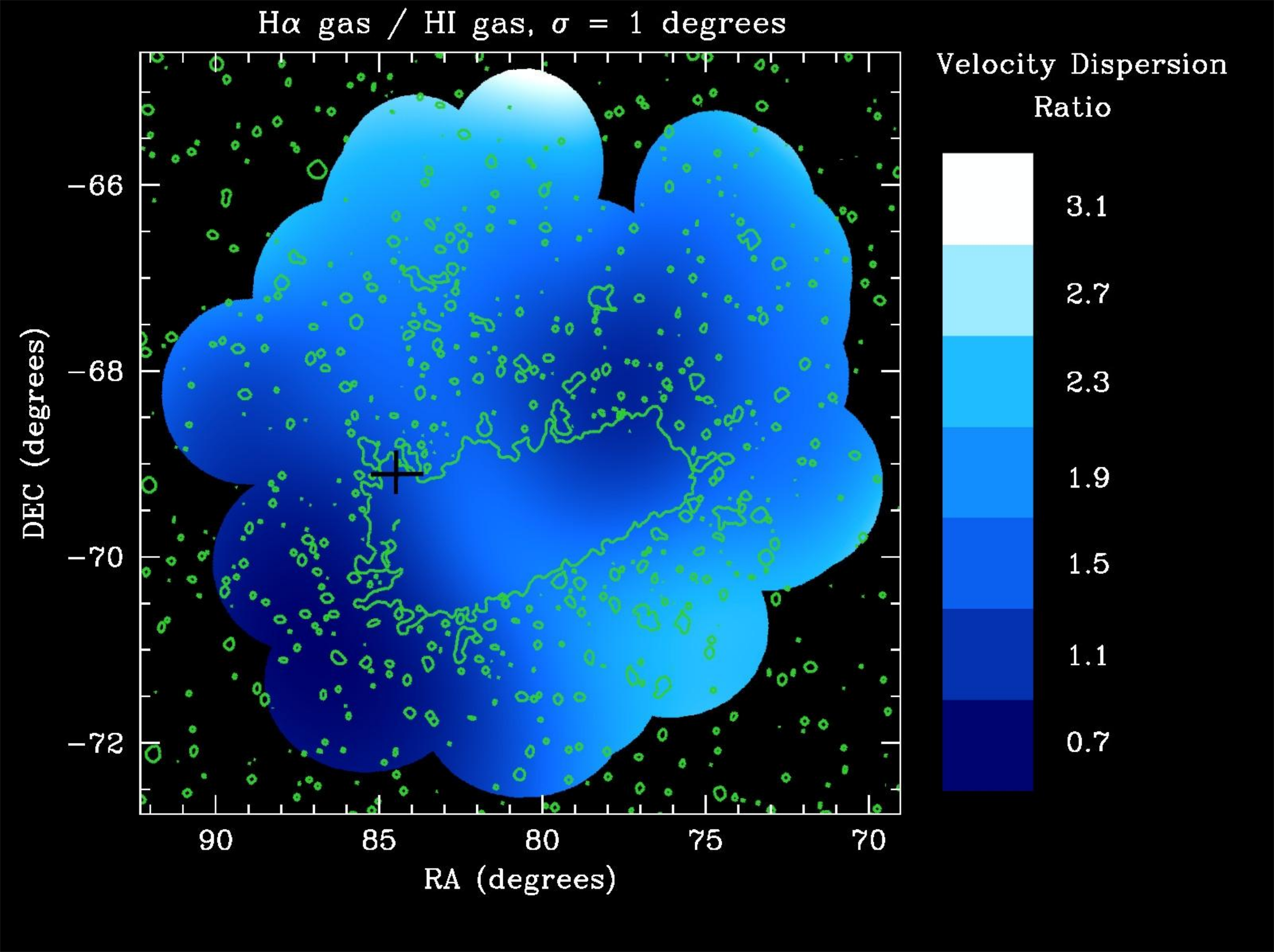}
  \caption{The top panel is the interpolation across the LMC of the velocity dispersion derived from the \Halpha\ emission line using a Gaussian of $\sigma = 1$\,deg; the interpolation has a cutoff distance from the nearest field equal to the $\sigma$ value. The middle panel is similar, except it shows the \HI\ velocity dispersion smoothed to $\sigma = 1$\,deg over the same region as the top panel. The bottom panel shows the ratio of the \Halpha\ velocity dispersion map divided by \HI\ map. The figures are resampled so they are linear in right ascension and declination. The contours are from the SHASSA Continuum Image and highlight the LMC stellar bar. The black cross in the panels is the location of 30 Doradus.}
  \label{fig:velocity_dispersion}
\end{figure}

We create interpolated \Halpha\ velocity dispersion maps the same as for radial velocity. The top panel of Figure~\ref{fig:velocity_dispersion_focussed} shows an interpolated map of the velocity dispersion created with $\sigma = 0.15$\,deg, while the top panel of Figure \ref{fig:velocity_dispersion} shows an interpolated map with $\sigma = 1$\,deg. The region around 30~Doradus (at RA=84.68\degrees, Dec=$-$69.10\degrees) has the highest velocity dispersion, likely caused by the higher star formation rate exciting the gas there. Constellation~III, the region with the highest gas phase metallicity, is not obviously unusual in the velocity dispersion maps.

The raw \HI\ velocity dispersion (moment~2) map from \cite{kim03} and \citet{staveleysmith03} is shown in the bottom panel of Figure~\ref{fig:HI_raw} at a resolution of 1~arcmin for reference. The velocity dispersion maps of the \HI\ gas from can be seen in the middle panel of Figure \ref{fig:velocity_dispersion_focussed} with smoothing $\sigma = 0.15$\,deg and in the middle panel of Figure \ref{fig:velocity_dispersion} with smoothing $\sigma = 1$\,deg. These maps have been limited to the same area covered in our \Halpha\ velocity dispersion maps for ease of comparison.

The ratio between the \Halpha\ and \HI\ velocity dispersion maps can be seen in the bottom panels of Figure~\ref{fig:velocity_dispersion_focussed} and Figure~\ref{fig:velocity_dispersion}, with smoothing $\sigma = 0.15$\,deg and $\sigma = 1$\,deg respectively. The peak region close to 30~Doradus (at RA=84.68\degrees, Dec=-69.10\degrees) in the \HI\ velocity dispersion maps shows a ratio close to 1 in the velocity dispersion ratio maps, indicating that this star formation region is affecting both \Halpha\ emitting gas and \HI\ gas equally.


While generally the \Halpha\ velocity dispersion is higher than the \HI\ velocity dispersion, there is a region in the south-east where the \HI\ velocity dispersion is higher. These values are too large to be attributed to thermal motions in the neutral gas.  A closer inspection of the velocity distribution in this region shows no evidence for multiple peaks, thus we interpret this heightened dispersion as turbulence on small scales.  This region lies just below 30~Doradus. This is the same region where the \Halpha\ radial velocity and \HI\ radial velocity diverged. As it is the \HI\ gas in this region that has the higher velocity dispersion, it is likely that the radial velocity difference is due to a divergence in the \HI\ gas as well, i.e.\ the \Halpha\ emitting gas tracks the rotation of the LMC disk while the \HI\ gas is red-shifted from the bulk motion and has higher velocity dispersion. This could be an outflow leaving from the other side of the LMC or an inflow from this side of the LMC. If an inflow, then the gas is likely moving towards the nearby 30~Doradus, possibly acting as fuel for the star formation there. In the higher resolution \HI\ observations, this region stands out clearly in both radial velocity and velocity dispersion. This feature does not lie in the direction of the Magellanic Stream and is a much more localised phenomenon.


\section{Conclusion}
\label{Conclusion}

The properties of optical emission lines in many \HII\ regions across the Large Magellanic Cloud have been measured using bright-time observations with the WiFeS instrument on the ANU 2.3-metre telescope. These measurements allow us to derive the gas phase metallicity in multiple fields and to interpolate the most complete map to date of the gas phase metallicity across the LMC. This map shows that there is no simple gradient (radial or otherwise), but rather a complex distribution that is consistent with the irregular nature of the LMC. The region with the highest gas phase metallicity lies in the north of the LMC. 

We have also derived the extinction in multiple fields across the LMC from the flux ratio of the \Halpha\ and \Hbeta\ emission lines. Interpolating a map across the LMC, 30~Doradus stands out as a significant region of extinction.

We have compared measurements of the radial velocity within the LMC made from our \Halpha\ emission line measurements and from \HI\ 21~cm emission measurements. The velocities mostly agree well, except for a region to the east and south, just below 30~Doradus, where the \Halpha\ and \HI\ gas radial velocities diverge. We have also compared measurements of the velocity dispersion from \Halpha\ and \HI. 30~Doradus shows up as having the highest \Halpha\ emission line velocity dispersion, indicating that it has the highest levels of ionized gas turbulence in the LMC. 

In general, the \Halpha\ velocity dispersion is higher than the \HI\ velocity dispersion. The velocity dispersion difference between \Halpha\ emitting gas and \HI\ gas is near zero around 30~Doradus, showing that this region strongly affects both hot and cold gas. The maximum difference is where the \HI\ gas has the minimum velocity dispersion. There is a region to the south and east where the \HI\ velocity dispersion is larger than the \Halpha\ velocity dispersion; this is the same region where the radial velocities diverge. This suggests the \HI\ gas is diverging from the bulk motion of the stars, indicating either an outflow or inflow of \HI\ gas in this region.

{The maps of the LMC presented here show that dwarf galaxies can have a complex structure that might be missed when observing them at large distances.  This complex structure is important, as it may affect the formation of larger galaxies when these smaller systems merge.}   


\section*{Acknowledgements}

We wish to acknowledge the help of Naomi McClure-Griffiths, Lister Staveley-Smith, Mike Bessell, Ian Price, Christopher Lidman, Jon Nielsen, Chris Dowling, and Se-Heon Oh with this work. MMC acknowledges support from a Royal Society Wolfson Visiting Fellowship (RSWVF{\textbackslash}R3{\textbackslash}223005) while on sabbatical at the University of Oxford. 
FDE acknowledges support by the Science and Technology Facilities Council (STFC), by the ERC through Advanced Grant 695671 ``QUENCH'', and by the UKRI Frontier Research grant RISEandFALL. We make use of data from the Southern H-Alpha Sky Survey Atlas (SHASSA), which is supported by the National Science Foundation.


\section*{Data Availability}

The raw spectrographic data used for this project is stored in the WiFeS archive. Reduced spectrographic data is available by contacting the first author. \HI\ data is available on the ATNF HI Survey Databases website, found here https://www.atnf.csiro.au/observers/data/data\_HI.html. SHASSA H$\alpha$\ emission survey data can be found at http://amundsen.swarthmore.edu/SHASSA.



\bibliographystyle{mnras}
\bibliography{LMC.bib} 


\section*{Appendix}

The Appendix contains tables of the values for each field used in this work.


\begin{table*}
\centering
\caption{The 83 WiFeS fields, with their \Halpha\ fluxes, radial velocities and velocity dispersions. The R.A.\ and Dec.\ are in degrees, the fluxes are in erg/s/cm$^2$/\AA, and the radial velocities and velocity dispersions are in \kms. The error in the radial velocity is the random error; there is a systematic error from the wavelength correction that can be much larger, varying from 0 to 3~\kms.  The error on the velocity dispersion is the random error.  Based on the sky lines there is an error \around 1.7~\kms\ on the grating resolution in this region. The table continues in Table \ref{tab:Halpha_2}.}
\label{tab:Halpha}
\begin{tabular}{ccccclcl} 
\hline
     &      &              &       & \Halpha\ radial &       & \Halpha\ velocity &       \\ 
R.A. & Dec. & \Halpha\ flux & error & velocity        & error & dispersion        & error \\ 
\hline
  72.3592 &  -69.1992 & 3.747e-11 &   1.5e-13 &    252.24 &     0.089 &     10.82 &      0.12 \\ 
  73.0296 &  -66.9250 & 5.915e-11 &   2.7e-13 &    286.74 &     0.099 &     15.56 &      0.12 \\ 
  73.1075 &  -69.3692 & 7.030e-12 &   6.9e-14 &    251.63 &      0.24 &     10.58 &      0.31 \\ 
  73.1425 &  -67.2811 & 4.992e-12 &   3.8e-14 &    290.73 &      0.14 &     10.27 &      0.24 \\ 
  73.2583 &  -68.0497 & 2.022e-11 &   1.3e-13 &    284.85 &      0.13 &     11.85 &      0.16 \\ 
  73.5088 &  -69.2006 & 1.111e-10 &   2.9e-13 &    260.20 &     0.052 &     13.31 &      0.06 \\ 
  73.7779 &  -67.1878 & 4.502e-13 &   1.5e-14 &    297.12 &       1.5 &     21.83 &       1.5 \\ 
  74.0675 &  -66.3103 & 5.035e-13 &   2.6e-14 &    296.24 &       1.4 &     19.21 &       1.8 \\ 
  74.1983 &  -66.4058 & 3.694e-11 &   1.9e-13 &    298.66 &      0.13 &     12.19 &      0.15 \\ 
  74.2325 &  -66.5969 & 1.682e-12 &   5.4e-14 &    289.20 &      0.71 &     11.61 &      0.85 \\ 
  74.2492 &  -68.7506 & 6.642e-12 &   3.6e-14 &    281.10 &      0.11 &     11.92 &      0.19 \\ 
  74.3258 &  -68.4028 & 6.172e-12 &   6.8e-14 &    273.58 &      0.39 &     18.14 &      0.48 \\ 
  74.4142 &  -66.4558 & 3.376e-11 &   9.9e-14 &    294.20 &      0.06 &     10.87 &     0.061 \\ 
  74.6967 &  -66.1958 & 2.160e-12 &   2.8e-14 &    297.43 &      0.28 &     8.572 &      0.33 \\ 
  76.0863 &  -70.7328 & 8.229e-12 &   5.8e-14 &    242.51 &       0.2 &     11.06 &      0.21 \\ 
  76.2579 &  -68.0567 & 7.948e-12 &   6.0e-14 &    276.00 &      0.17 &     9.997 &      0.21 \\ 
  76.5167 &  -68.1144 & 9.823e-13 &   2.3e-14 &    266.56 &      0.56 &     8.549 &      0.97 \\ 
  78.3667 &  -69.3403 & 1.142e-12 &   6.8e-14 &    255.36 &       1.2 &     10.96 &       1.7 \\ 
  78.4375 &  -67.4347 & 3.397e-13 &   9.9e-15 &    309.19 &      0.68 &     17.25 &       1.1 \\ 
  78.5617 &  -67.1428 & 1.410e-12 &   3.7e-14 &    303.01 &      0.71 &     13.05 &       1.2 \\ 
  79.2021 &  -69.2075 & 7.807e-13 &   3.5e-14 &    272.37 &       1.1 &     12.42 &       1.6 \\ 
  79.4246 &  -71.2628 & 1.002e-11 &   7.1e-14 &    236.06 &      0.17 &     13.33 &      0.27 \\ 
  79.4975 &  -69.0975 & 3.969e-12 &   4.0e-14 &    268.35 &      0.19 &     13.10 &      0.25 \\ 
  79.7708 &  -69.1919 & 1.151e-11 &   6.9e-14 &    286.71 &      0.16 &     10.40 &      0.17 \\ 
  80.5504 &  -65.7433 & 2.364e-12 &   5.0e-14 &    312.74 &      0.44 &     11.84 &      0.61 \\ 
  80.5579 &  -71.5961 & 1.526e-12 &   3.1e-14 &    236.84 &      0.54 &     11.27 &      0.59 \\ 
  80.6471 &  -68.0375 & 1.399e-12 &   2.1e-14 &    293.05 &      0.46 &     10.76 &      0.52 \\ 
  80.8962 &  -71.3272 & 9.836e-13 &   2.1e-14 &    230.78 &      0.64 &     10.76 &      0.93 \\ 
  81.1654 &  -68.5011 & 8.062e-13 &   4.1e-14 &    269.94 &       1.1 &     13.22 &       1.3 \\ 
  81.3608 &  -69.2183 & 8.594e-13 &   3.3e-14 &    269.78 &      0.85 &     15.12 &       1.4 \\ 
  81.5017 &  -66.2644 & 1.659e-12 &   4.3e-14 &    305.10 &      0.63 &     12.69 &      0.82 \\ 
  81.5225 &  -68.6294 & 1.109e-12 &   4.5e-14 &    271.21 &       1.1 &     11.56 &       1.6 \\ 
  81.5654 &  -68.9069 & 1.482e-12 &   5.0e-14 &    268.30 &       1.0 &     21.53 &       1.6 \\ 
  81.6075 &  -67.4725 & 1.393e-11 &   1.0e-13 &    309.77 &      0.18 &     12.73 &      0.21 \\ 
  81.6246 &  -67.6108 & 6.351e-12 &   9.3e-14 &    305.82 &      0.34 &     11.54 &      0.46 \\ 
  81.6296 &  -69.3150 & 2.804e-12 &   4.9e-14 &    278.52 &      0.42 &     11.94 &      0.49 \\ 
  81.8188 &  -70.5839 & 1.395e-12 &   3.0e-14 &    241.48 &      0.50 &     11.91 &      0.78 \\ 
  81.9571 &  -67.4250 & 4.276e-12 &   7.3e-14 &    311.88 &      0.60 &     13.09 &      0.65 \\ 
  82.1583 &  -69.0217 & 1.284e-12 &   3.8e-14 &    264.71 &      0.95 &     19.80 &       1.3 \\ 
  82.5083 &  -69.3722 & 6.454e-13 &   3.7e-14 &    262.57 &       1.3 &     14.08 &       1.9 \\ 
  82.5771 &  -68.6469 & 9.355e-13 &   3.2e-14 &    268.62 &       1.1 &     14.97 &       1.6 \\ 
  82.7358 &  -69.0569 & 1.425e-12 &   5.4e-14 &    265.50 &       1.3 &     20.48 &       1.7 \\ 
  82.8492 &  -71.0700 & 2.509e-11 &   1.6e-13 &    240.98 &      0.15 &     12.09 &      0.18 \\ 
  82.8650 &  -68.6267 & 1.171e-12 &   4.1e-14 &    271.77 &      0.74 &     15.26 &       1.2 \\ 
  82.9600 &  -69.2892 & 1.122e-12 &   5.8e-14 &    256.04 &       1.6 &     26.93 &       2.7 \\ 
  82.9796 &  -69.1861 & 8.138e-13 &   3.4e-14 &    279.48 &       1.4 &     19.55 &       2.2 \\ 
  82.9817 &  -67.6881 & 5.463e-12 &   5.0e-14 &    302.68 &      0.45 &     31.50 &       0.4 \\ 
  83.0296 &  -68.4700 & 6.042e-13 &   1.1e-14 &    274.93 &      0.51 &     13.42 &      0.77 \\ 
  83.0458 &  -66.4403 & 4.332e-12 &   4.4e-14 &    307.83 &      0.24 &     9.704 &      0.27 \\ 
  83.2150 &  -67.5792 & 1.335e-12 &   4.3e-14 &    303.31 &       1.0 &     19.62 &       1.3 \\ 
  83.2188 &  -68.9217 & 5.079e-13 &   2.0e-14 &    270.83 &       1.4 &     22.48 &       2.3 \\ 
  83.4746 &  -69.4017 & 1.053e-12 &   3.2e-14 &    258.06 &      0.95 &     18.50 &       1.4 \\ 
  83.5650 &  -68.7683 & 6.346e-13 &   2.8e-14 &    270.77 &       1.3 &     26.85 &       2.5 \\ 
  83.6204 &  -69.0822 & 8.362e-13 &   4.9e-14 &    271.24 &       1.8 &     18.09 &       2.4 \\ 
  83.6350 &  -69.6681 & 1.294e-12 &   4.3e-14 &    268.86 &       1.2 &     23.29 &       1.6 \\ 
  83.6400 &  -69.7817 & 3.338e-13 &   1.6e-14 &    272.06 &       1.3 &     14.68 &       1.7 \\ 
  83.7438 &  -69.2058 & 2.824e-12 &   5.2e-14 &    262.27 &      0.59 &     21.00 &      0.90 \\ 
  83.8108 &  -68.9119 & 6.940e-13 &   2.9e-14 &    269.64 &       1.5 &     22.46 &       2.2 \\ 
\end{tabular}
\end{table*}

\begin{table*}
\centering
\caption{The continuance of Table \ref{tab:Halpha}.}
\label{tab:Halpha_2}
\begin{tabular}{ccccclcl} 
\hline
     &      &              &       & \Halpha\ radial &       & \Halpha\ velocity &       \\ 
R.A. & Dec. & \Halpha\ flux & error & velocity        & error & dispersion        & error \\ 
\hline
  83.8283 &  -67.5800 & 2.598e-11 &   9.7e-14 &    298.95 &     0.087 &     12.49 &      0.10 \\ 
  83.8608 &  -69.3097 & 1.112e-11 &   4.6e-14 &    266.53 &     0.097 &     17.40 &      0.14 \\ 
  83.8879 &  -66.0267 & 4.132e-12 &   9.7e-14 &    301.61 &      0.47 &     11.71 &      0.66 \\ 
  84.1671 &  -67.4581 & 6.436e-13 &   2.1e-14 &    298.87 &       1.1 &     15.57 &       1.3 \\ 
  84.2375 &  -69.0525 & 4.731e-12 &   2.8e-14 &    273.37 &      0.23 &     26.84 &      0.31 \\ 
  84.2979 &  -66.3133 & 1.423e-12 &   2.9e-14 &    305.96 &      0.52 &     9.846 &      0.75 \\ 
  84.3100 &  -69.2953 & 7.402e-12 &   7.2e-14 &    268.21 &      0.38 &     17.98 &      0.42 \\ 
  84.3992 &  -69.4564 & 1.253e-12 &   4.5e-14 &    285.27 &       1.5 &     29.99 &       1.8 \\ 
  84.4729 &  -69.1894 & 1.694e-11 &   7.0e-14 &    272.15 &      0.12 &     20.08 &       0.2 \\ 
  84.5883 &  -70.6858 & 3.544e-11 &   1.2e-13 &    227.89 &     0.074 &     12.58 &      0.08 \\ 
  84.7779 &  -69.6886 & 1.246e-12 &   3.1e-14 &    250.95 &      0.72 &     14.52 &       1.0 \\ 
  84.8546 &  -69.4556 & 6.477e-12 &   1.2e-13 &    265.78 &      0.57 &     19.29 &      0.65 \\ 
  84.8567 &  -69.3319 & 1.091e-12 &   3.6e-14 &    254.41 &       1.3 &     28.32 &       2.0 \\ 
  84.9592 &  -69.0281 & 3.706e-12 &   5.8e-14 &    266.07 &      0.63 &     22.36 &      0.81 \\ 
  85.0600 &  -69.1756 & 1.111e-11 &   9.9e-14 &    272.74 &      0.28 &     17.37 &      0.37 \\ 
  85.1975 &  -68.9608 & 9.524e-13 &   3.2e-14 &    263.26 &       1.1 &     19.06 &       1.4 \\ 
  85.2017 &  -69.7883 & 6.019e-13 &   1.7e-14 &    260.17 &      0.72 &     14.69 &       1.3 \\ 
  85.4029 &  -71.3242 & 1.331e-11 &   9.5e-14 &    230.24 &      0.16 &     10.62 &      0.16 \\ 
  85.5708 &  -69.0381 & 9.093e-13 &   2.1e-14 &    269.56 &      0.75 &     19.00 &       1.0 \\ 
  85.7767 &  -69.7506 & 5.961e-12 &   7.0e-14 &    248.69 &      0.26 &     12.04 &      0.37 \\ 
  85.8063 &  -67.8450 & 6.995e-13 &   1.4e-14 &    305.49 &      0.45 &     13.88 &      0.61 \\ 
  86.3108 &  -67.1556 & 2.668e-12 &   4.7e-14 &    299.60 &      0.43 &     10.08 &      0.54 \\ 
  86.3513 &  -69.7769 & 8.720e-12 &   1.2e-13 &    237.67 &      0.32 &     9.916 &      0.44 \\ 
  87.1817 &  -70.0469 & 2.761e-12 &   6.8e-14 &    241.52 &      0.72 &     9.308 &      0.74 \\ 
  88.9846 &  -68.2239 & 3.657e-13 &   1.4e-14 &    302.50 &      0.88 &     10.02 &       1.1 \\ 
\end{tabular}
\end{table*}


\begin{table*}
\centering
\caption{The 59 WiFeS fields with their line fluxes and metallicity values. The R.A. and Dec. are in degrees, the fluxes are in erg\,s$^{-1}$\,cm$^{-2}$\,\AA$^{-1}$, and the metallicities are in dex.}
\label{tab:metallicity}
\begin{tabular}{cccccccccccl} 
\hline
     &      & \Hbeta &       & [OIII]5007$\lambda$  &       & \Halpha &       & [NII]6584$\lambda$  &       & O3N2        &       \\
R.A. & Dec. & flux   & error & flux                 & error & flux    & error & flux                & error & metallicity & error \\
\hline
  72.3592 &  -69.1992 & 4.460e-12 &   1.0e-13 & 1.195e-11 &   1.1e-13 & 3.747e-11 &   1.5e-13 & 2.630e-12 &   5.5e-14 &    8.2239 &    0.0045 \\ 
  73.0296 &  -66.9250 & 1.640e-11 &   2.1e-13 & 6.142e-11 &   3.2e-13 & 5.915e-11 &   2.7e-13 & 2.865e-12 &   6.0e-14 &    8.1258 &    0.0035 \\ 
  73.1075 &  -69.3692 & 1.977e-12 &   7.1e-14 & 3.543e-12 &   7.1e-14 & 7.030e-12 &   6.9e-14 & 8.250e-13 &   3.5e-14 &    8.3512 &    0.0083 \\ 
  73.1425 &  -67.2811 & 1.272e-12 &   4.8e-14 & 1.869e-12 &   6.5e-14 & 4.992e-12 &   3.8e-14 & 5.461e-13 &   2.4e-14 &    8.3690 &    0.0094 \\ 
  73.2583 &  -68.0497 & 6.361e-12 &   1.4e-13 & 2.153e-11 &   1.7e-13 & 2.022e-11 &   1.3e-13 & 8.483e-13 &   3.7e-14 &    8.1198 &    0.0070 \\ 
  73.5088 &  -69.2006 & 1.691e-10 &   6.0e-13 & 6.475e-10 &   8.9e-13 & 1.111e-10 &   2.9e-13 & 4.353e-12 &   6.4e-14 &    8.0932 &    0.0021 \\ 
  73.7779 &  -67.1878 & 1.438e-13 &   2.1e-14 & 2.782e-13 &   2.5e-14 & 4.502e-13 &   1.5e-14 & 6.195e-14 &   1.2e-14 &    8.3627 &     0.036 \\ 
  74.1983 &  -66.4058 & 1.056e-11 &   1.6e-13 & 3.259e-11 &   2.6e-13 & 3.694e-11 &   1.9e-13 & 1.976e-12 &   5.4e-14 &    8.1664 &    0.0046 \\ 
  74.4142 &  -66.4558 & 9.264e-12 &   7.5e-14 & 2.943e-11 &   8.1e-14 & 3.376e-11 &   9.9e-14 & 1.972e-12 &   2.5e-14 &    8.1747 &    0.0022 \\ 
  74.6967 &  -66.1958 & 6.118e-13 &   3.2e-14 & 7.187e-13 &   2.8e-14 & 2.160e-12 &   2.8e-14 & 1.840e-13 &   1.5e-14 &    8.3654 &     0.015 \\ 
  76.0863 &  -70.7328 & 2.398e-12 &   8.2e-14 & 4.430e-12 &   6.9e-14 & 8.229e-12 &   5.8e-14 & 6.061e-13 &   3.9e-14 &    8.2822 &     0.010 \\ 
  76.2579 &  -68.0567 & 2.275e-12 &   7.3e-14 & 6.160e-12 &   8.6e-14 & 7.948e-12 &   6e.0-14 & 4.188e-13 &   2.4e-14 &    8.1825 &    0.0095 \\ 
  78.4375 &  -67.4347 & 1.658e-13 &   1.7e-14 & 7.009e-13 &   1.8e-14 & 3.397e-13 &   9.9e-15 & 2.931e-14 &   7.4e-15 &    8.1892 &     0.038 \\ 
  79.4246 &  -71.2628 & 2.714e-12 &   7.9e-14 & 7.473e-12 &   8.8e-14 & 1.002e-11 &   7.1e-14 & 5.214e-13 &   3.1e-14 &    8.1785 &    0.0094 \\ 
  79.6867 &  -69.6631 & 1.033e-12 &   8.4e-14 & 1.409e-12 &   8.1e-14 & 3.415e-12 &   8.2e-14 & 5.563e-13 &   5.1e-14 &    8.4346 &     0.019 \\ 
  79.7708 &  -69.1919 & 2.015e-11 &   1.7e-13 & 2.722e-11 &   1.5e-13 & 1.151e-11 &   6.9e-14 & 1.513e-12 &   3.2e-14 &    8.4062 &    0.0033 \\ 
  80.5504 &  -65.7433 & 2.394e-13 &   4.5e-14 & 4.249e-13 &   3.4e-14 & 2.364e-12 &   5.0e-14 & 3.241e-13 &   2.8e-14 &    8.3741 &     0.031 \\ 
  80.5517 &  -67.9025 & 1.760e-12 &   7.5e-14 & 1.274e-12 &   4.7e-14 & 5.692e-12 &   6.9e-14 & 7.338e-13 &   3.1e-14 &    8.4902 &    0.0099 \\ 
  80.5579 &  -71.5961 & 4.699e-13 &   5.3e-14 & 1.538e-12 &   5.5e-14 & 1.526e-12 &   3.1e-14 & 1.186e-13 &   2.1e-14 &    8.2103 &     0.030 \\ 
  81.3608 &  -69.2183 & 2.552e-13 &   4.5e-14 & 4.747e-13 &   4.4e-14 & 8.594e-13 &   3.3e-14 & 1.442e-13 &   2.7e-14 &    8.3956 &     0.038 \\ 
  81.5017 &  -66.2644 & 4.727e-13 &   4.7e-14 & 4.001e-13 &   3.6e-14 & 1.659e-12 &   4.3e-14 & 2.037e-13 &   2.6e-14 &    8.4617 &     0.026 \\ 
  81.5654 &  -68.9069 & 3.415e-13 &   5.4e-14 & 1.054e-12 &   1.0e-13 & 1.482e-12 &   5.0e-14 & 2.029e-13 &   3.6e-14 &    8.2969 &     0.036 \\ 
  81.6075 &  -67.4725 & 2.304e-11 &   2.8e-13 & 3.087e-11 &   2.3e-13 & 1.393e-11 &   1.0e-13 & 1.470e-12 &   5.0e-14 &    8.3768 &    0.0053 \\ 
  81.6246 &  -67.6108 & 2.239e-12 &   1.5e-13 & 4.205e-12 &   1.4e-13 & 6.351e-12 &   9.3e-14 & 5.251e-13 &   5.4e-14 &    8.2960 &     0.018 \\ 
  81.6296 &  -69.3150 & 9.107e-13 &   6.7e-14 & 1.971e-12 &   5.9e-14 & 2.804e-12 &   4.9e-14 & 1.818e-13 &   2.9e-14 &    8.2425 &     0.025 \\ 
  81.9571 &  -67.4250 & 6.021e-13 &   1.2e-13 & 4.127e-13 &   7.0e-14 & 4.276e-12 &   7.3e-14 & 6.164e-13 &   5.9e-14 &    8.5133 &     0.039 \\ 
  82.1583 &  -69.0217 & 3.815e-13 &   6.0e-14 & 3.623e-13 &   5.9e-14 & 1.284e-12 &   3.8e-14 & 2.180e-13 &   2.4e-14 &    8.4908 &     0.035 \\ 
  82.5771 &  -68.6469 & 2.725e-13 &   3.8e-14 & 3.994e-13 &   5.5e-14 & 9.355e-13 &   3.2e-14 & 1.580e-13 &   3.3e-14 &    8.4298 &     0.040 \\ 
  82.7358 &  -69.0569 & 4.123e-13 &   9.1e-14 & 7.037e-13 &   7.7e-14 & 1.425e-12 &   5.4e-14 & 2.385e-13 &   6.6e-14 &    8.4073 &     0.052 \\ 
  82.8492 &  -71.0700 & 5.796e-12 &   1.2e-13 & 1.153e-11 &   1.3e-13 & 2.509e-11 &   1.6e-13 & 2.229e-12 &   5.4e-14 &    8.2979 &    0.0047 \\ 
  82.8650 &  -68.6267 & 3.924e-13 &   6.3e-14 & 4.279e-13 &   5.0e-14 & 1.171e-12 &   4.1e-14 & 1.704e-13 &   2.0e-14 &    8.4501 &     0.032 \\ 
  82.9600 &  -69.2892 & 3.196e-13 &   7.8e-14 & 7.186e-13 &   8.1e-14 & 1.122e-12 &   5.8e-14 & 1.682e-13 &   4.7e-14 &    8.3537 &     0.054 \\ 
  82.9817 &  -67.6881 & 9.322e-12 &   2.1e-13 & 1.356e-11 &   1.7e-13 & 5.463e-12 &   5.0e-14 & 7.646e-13 &   3.6e-14 &    8.4047 &    0.0075 \\ 
  83.0458 &  -66.4403 & 1.274e-12 &   4.7e-14 & 3.652e-13 &   3.5e-14 & 4.332e-12 &   4.4e-14 & 5.423e-13 &   2.2e-14 &    8.6148 &     0.015 \\ 
  83.2150 &  -67.5792 & 3.961e-13 &   6.6e-14 & 5.463e-13 &   7.3e-14 & 1.335e-12 &   4.3e-14 & 1.783e-13 &   4.9e-14 &    8.4056 &     0.048 \\ 
  83.2188 &  -68.9217 & 1.612e-13 &   3.2e-14 & 2.881e-13 &   4.0e-14 & 5.079e-13 &   2.0e-14 & 7.786e-14 &   1.8e-14 &    8.3886 &     0.047 \\ 
  83.3883 &  -67.4600 & 5.829e-13 &   7.7e-14 & 1.203e-12 &   6.7e-14 & 1.899e-12 &   4.1e-13 & 2.616e-13 &   4.5e-14 &    8.3539 &     0.043 \\ 
  83.5650 &  -68.7683 & 2.044e-13 &   2.8e-14 & 4.996e-13 &   2.7e-14 & 6.346e-13 &   2.8e-14 & 8.393e-14 &   1.9e-14 &    8.3247 &     0.038 \\ 
  83.6350 &  -69.6681 & 3.750e-13 &   6.2e-14 & 7.368e-13 &   5.6e-14 & 1.294e-12 &   4.3e-14 & 1.765e-13 &   4.0e-14 &    8.3592 &     0.041 \\ 
  83.6400 &  -69.7817 & 1.035e-13 &   2.3e-14 & 2.069e-13 &   2.0e-14 & 3.338e-13 &   1.6e-14 & 5.076e-14 &   1.3e-14 &    8.3720 &     0.050 \\ 
  83.7438 &  -69.2058 & 8.600e-13 &   6.4e-14 & 2.757e-12 &   7.3e-14 & 2.824e-12 &   5.2e-14 & 1.869e-13 &   3.1e-14 &    8.1907 &     0.025 \\ 
  83.8283 &  -67.5800 & 7.867e-12 &   7.5e-14 & 3.538e-11 &   1.2e-13 & 2.598e-11 &   9.7e-14 & 9.512e-13 &   2.2e-14 &    8.0614 &    0.0036 \\ 
  83.8608 &  -69.3097 & 6.841e-12 &   7.8e-14 & 1.346e-11 &   7.2e-14 & 1.112e-11 &   4.6e-14 & 9.740e-13 &   2.3e-14 &    8.2975 &    0.0038 \\ 
  84.2375 &  -69.0525 & 1.274e-12 &   2.6e-14 & 4.285e-12 &   2.9e-14 & 4.731e-12 &   2.8e-14 & 3.851e-13 &   1.9e-14 &    8.2128 &    0.0073 \\ 
  84.3992 &  -69.4564 & 3.414e-13 &   8.6e-14 & 7.125e-13 &   6.2e-14 & 1.253e-12 &   4.5e-14 & 1.828e-13 &   3.9e-14 &    8.3602 &     0.048 \\ 
  84.4729 &  -69.1894 & 3.934e-12 &   4.8e-14 & 1.100e-11 &   5.3e-14 & 1.694e-11 &   7.0e-14 & 1.489e-12 &   2.3e-14 &    8.2492 &    0.0028 \\ 
  84.5883 &  -70.6858 & 9.086e-12 &   7.6e-14 & 3.563e-11 &   1.1e-13 & 3.544e-11 &   1.2e-13 & 1.826e-12 &   2.5e-14 &    8.1279 &    0.0023 \\ 
  84.6575 &  -69.0914 & 5.982e-12 &   1.2e-13 & 3.222e-11 &   2.4e-13 & 1.377e-11 &   1.2e-13 & 7.325e-13 &   2.9e-14 &    8.0883 &    0.0063 \\ 
  84.8546 &  -69.4556 & 1.011e-11 &   2.4e-13 & 1.777e-11 &   2.4e-13 & 6.477e-12 &   1.2e-13 & 1.015e-12 &   6.4e-14 &    8.3939 &    0.0099 \\ 
  84.8567 &  -69.3319 & 3.191e-13 &   5.9e-14 & 6.008e-13 &   5.5e-14 & 1.091e-12 &   3.6e-14 & 1.522e-13 &   3.3e-14 &    8.3683 &     0.042 \\ 
  84.9592 &  -69.0281 & 5.977e-12 &   2.3e-13 & 1.615e-11 &   2.3e-13 & 3.706e-12 &   5.8e-14 & 4.024e-13 &   3.4e-14 &    8.2833 &     0.013 \\ 
  85.0600 &  -69.1756 & 1.223e-12 &   9.4e-14 & 4.494e-12 &   9.0e-14 & 1.111e-11 &   9.9e-14 & 7.395e-13 &   5.5e-14 &    8.1725 &     0.015 \\ 
  85.1975 &  -68.9608 & 3.025e-13 &   7.2e-14 & 6.112e-13 &   6.1e-14 & 9.524e-13 &   3.2e-14 & 1.378e-13 &   2.5e-14 &    8.3636 &     0.044 \\ 
  85.2017 &  -69.7883 & 1.629e-13 &   1.7e-14 & 2.491e-13 &   1.9e-14 & 6.019e-13 &   1.7e-14 & 8.577e-14 &   1.1e-14 &    8.4002 &     0.025 \\ 
  85.3229 &  -69.6317 & 2.842e-13 &   8.1e-14 & 3.640e-13 &   6.8e-14 & 9.513e-13 &   6.5e-14 & 1.473e-13 &   4.9e-14 &    8.4364 &     0.067 \\ 
  85.4029 &  -71.3242 & 4.007e-12 &   9.0e-14 & 1.349e-11 &   8.1e-14 & 1.331e-11 &   9.5e-14 & 7.415e-13 &   2.9e-14 &    8.1600 &    0.0064 \\ 
  85.5708 &  -69.0381 & 2.743e-13 &   1.9e-14 & 5.568e-13 &   2.0e-14 & 9.093e-13 &   2.1e-14 & 1.361e-13 &   1.3e-14 &    8.3677 &     0.018 \\ 
  85.8063 &  -67.8450 & 3.862e-13 &   1.5e-14 & 6.848e-13 &   1.4e-14 & 6.995e-13 &   1.4e-14 & 8.916e-14 &   7.5e-15 &    8.3641 &     0.013 \\ 
  86.3108 &  -67.1556 & 8.406e-13 &   6.6e-14 & 2.578e-12 &   6.1e-14 & 2.668e-12 &   4.7e-14 & 2.294e-13 &   2.5e-14 &    8.2333 &     0.019 \\ 
\end{tabular}
\end{table*}
 

\begin{table*}
\centering
\caption{The 28 WiFeS fields with their line fluxes and extinction values.  The RA and DEC are in degrees, the fluxes are in erg/s/cm$^2$/\AA, and the extinctions are in magnitudes.}
\label{tab:EBV}
\begin{tabular}{ccccccll} 
\hline
R.A. & Dec. & \Hbeta\ flux & error & \Halpha\ flux & error & $\rm E(B-V)_{H\beta - H\alpha}$ & error \\ 
\hline

  73.0296 &  -66.9250 & 1.640e-11 &   2.1e-13 & 5.915e-11 &   2.7e-13 &    0.1785 &     0.010 \\ 
  73.1075 &  -69.3692 & 1.977e-12 &   7.1e-14 & 7.030e-12 &   6.9e-14 &    0.1675 &     0.028 \\ 
  73.1425 &  -67.2811 & 1.272e-12 &   4.8e-14 & 4.992e-12 &   3.8e-14 &    0.2437 &     0.029 \\ 
  73.2583 &  -68.0497 & 6.361e-12 &   1.4e-13 & 2.022e-11 &   1.3e-13 &   0.08122 &     0.017 \\ 
  74.1983 &  -66.4058 & 1.056e-11 &   1.6e-13 & 3.694e-11 &   1.9e-13 &    0.1553 &     0.012 \\ 
  74.3258 &  -68.4028 & 1.878e-12 &   7.2e-14 & 6.172e-12 &   6.8e-14 &    0.1072 &     0.030 \\ 
  74.4142 &  -66.4558 & 9.264e-12 &   7.5e-14 & 3.376e-11 &   9.9e-14 &    0.1867 &    0.0063 \\ 
  74.6967 &  -66.1958 & 6.118e-13 &   3.2e-14 & 2.160e-12 &   2.8e-14 &    0.1622 &     0.041 \\ 
  76.0863 &  -70.7328 & 2.398e-12 &   8.2e-14 & 8.229e-12 &   5.8e-14 &    0.1403 &     0.026 \\ 
  76.2579 &  -68.0567 & 2.275e-12 &   7.3e-14 & 7.948e-12 &   6.0e-14 &    0.1543 &     0.025 \\ 
  79.4246 &  -71.2628 & 2.714e-12 &   7.9e-14 & 1.002e-11 &   7.1e-14 &    0.1963 &     0.022 \\ 
  79.6867 &  -69.6631 & 1.033e-12 &   8.4e-14 & 3.415e-12 &   8.2e-14 &    0.1118 &     0.063 \\ 
  80.5517 &  -67.9025 & 1.760e-12 &   7.5e-14 & 5.692e-12 &   6.9e-14 &   0.09453 &     0.033 \\ 
  80.5579 &  -71.5961 & 4.699e-13 &   5.3e-14 & 1.526e-12 &   3.1e-14 &   0.09758 &     0.088 \\ 
  81.5017 &  -66.2644 & 4.727e-13 &   4.7e-14 & 1.659e-12 &   4.3e-14 &    0.1576 &     0.076 \\ 
  81.6296 &  -69.3150 & 9.107e-13 &   6.7e-14 & 2.804e-12 &   4.9e-14 &   0.05687 &     0.057 \\ 
  82.8492 &  -71.0700 & 5.796e-12 &   1.2e-13 & 2.509e-11 &   1.6e-13 &    0.3192 &     0.016 \\ 
  83.0458 &  -66.4403 & 1.274e-12 &   4.7e-14 & 4.332e-12 &   4.4e-14 &    0.1335 &     0.028 \\ 
  83.7438 &  -69.2058 & 8.600e-13 &   6.4e-14 & 2.824e-12 &   5.2e-14 &    0.1064 &     0.057 \\ 
  83.8283 &  -67.5800 & 7.867e-12 &   7.5e-14 & 2.598e-11 &   9.7e-14 &    0.1107 &    0.0074 \\ 
  84.2375 &  -69.0525 & 1.274e-12 &   2.6e-14 & 4.731e-12 &   2.8e-14 &    0.2013 &     0.016 \\ 
  84.4729 &  -69.1894 & 3.934e-12 &   4.8e-14 & 1.694e-11 &   7.0e-14 &    0.3149 &    0.0093 \\ 
  84.5883 &  -70.6858 & 9.086e-12 &   7.6e-14 & 3.544e-11 &   1.2e-13 &     0.239 &    0.0065 \\ 
  85.2017 &  -69.7883 & 1.629e-13 &   1.7e-14 & 6.019e-13 &   1.7e-14 &    0.1974 &     0.081 \\ 
  85.4029 &  -71.3242 & 4.007e-12 &   9.0e-14 & 1.331e-11 &   9.5e-14 &    0.1153 &     0.017 \\ 
  85.5708 &  -69.0381 & 2.743e-13 &   1.9e-14 & 9.093e-13 &   2.1e-14 &    0.1136 &     0.053 \\ 
  86.3108 &  -67.1556 & 8.406e-13 &   6.6e-14 & 2.668e-12 &   4.7e-14 &   0.08026 &     0.061 \\ 
  86.3513 &  -69.7769 & 2.480e-12 &   1.3e-13 & 8.720e-12 &   1.2e-13 &     0.159 &     0.041 \\

\end{tabular}
\end{table*}


\bsp	
\label{lastpage}
\end{document}